\documentclass[letterpaper,twocolumn,10pt]{article}
\usepackage{usenix2019_v3}

\usepackage{algorithm}
\usepackage{algpseudocode}
\usepackage{amsmath}
\usepackage{amsfonts}
\usepackage{amsthm}
\usepackage{amssymb}
\usepackage{enumitem}
\usepackage{graphicx}
\usepackage{multirow}
\usepackage{subcaption}
\usepackage{textcomp}
\usepackage{todonotes}
\usepackage{comment}

\usepackage[subtle]{savetrees}

\usepackage{listings}
\newcommand{\oursystem}{Pollux}
\theoremstyle{definition}
\newtheorem{definition}{Definition}[section]

\newif\ifcommenton

\ifcommenton
\newcommand{\qirong}[1]{{\color{blue}{\bf\sf [Qirong: #1]}}}
\newcommand{\aurick}[1]{{\color{red}{\bf\sf [Aurick: #1]}}}
\newcommand{\williex}[1]{{\color{magenta}{\bf\sf [Willie: #1]}}}
\newcommand{\ericx}[1]{{\color{green}{\bf\sf [Eric: #1]}}}
\newcommand{\hao}[1]{{\color{orange}{\bf\sf [Hao: #1]}}}
\newcommand{\greg}[1]{{\color{cyan}{\bf\sf [Greg: #1]}}}
\else
\newcommand{\qirong}[1]{}
\newcommand{\aurick}[1]{}
\newcommand{\williex}[1]{}
\newcommand{\ericx}[1]{}
\newcommand{\hao}[1]{}
\newcommand{\greg}[1]{}
\fi

\usepackage[normalem]{ulem}

\usepackage[available,functional,reproduced]{usenixbadges}

\begin{document}



\title{
\Large \bf \oursystem{}: Co-adaptive Cluster Scheduling for Goodput-Optimized Deep Learning}

\author{
{\rm Aurick Qiao\textsuperscript{1,2}}
\qquad
{\rm Sang Keun Choe\textsuperscript{2}}
\qquad
{\rm Suhas Jayaram Subramanya\textsuperscript{2}}
\qquad
{\rm Willie Neiswanger\textsuperscript{1,2}}
\\
{\rm Qirong Ho\textsuperscript{1}}
\qquad
{\rm Hao Zhang\textsuperscript{1,3}}
\qquad
{\rm Gregory R. Ganger\textsuperscript{2}}
\qquad
{\rm Eric P. Xing\textsuperscript{4,1,2}}
\\
\\
\textsuperscript{1}Petuum, Inc.
\qquad
\textsuperscript{2}Carnegie Mellon University
\qquad
\textsuperscript{3}UC Berkeley
\qquad
\textsuperscript{4}MBZUAI
} 

\maketitle

\begin{abstract}
\oursystem{} improves scheduling performance in deep learning (DL) clusters by adaptively co-optimizing inter-dependent factors both at the per-job level and at the cluster-wide level. Most existing schedulers expect users to specify the number of resources for each job, often leading to inefficient resource use. Some recent schedulers choose job resources for users, but do so without awareness of how DL training can be re-optimized to better utilize the provided resources.

\oursystem{} simultaneously considers both aspects.
By monitoring the status of each job during training, \oursystem{} models how their \emph{goodput} (a metric we introduce to combine system throughput with statistical efficiency) would change by adding or removing resources.
\oursystem{} dynamically (re-)assigns resources to improve cluster-wide goodput, while respecting fairness and continually optimizing each DL job to better utilize those resources.

In experiments with real DL jobs and with trace-driven simulations, \oursystem{} reduces average job completion times by 37--50\% relative to state-of-the-art DL schedulers, even when they are provided with ideal resource and training configurations for every job.
\oursystem{} promotes fairness among DL jobs competing for resources,
based on a more meaningful measure of \textit{useful} job progress, 
and reveals a new opportunity for reducing DL cost in cloud environments.
\oursystem{} is implemented and publicly available as part of an open-source project at \url{https://github.com/petuum/adaptdl}.

\end{abstract}

\section{Introduction}

Deep learning (DL) training has rapidly become a dominant workload in many shared resource environments such as datacenters and the cloud.
DL jobs are resource-intensive and long-running, often demanding distributed execution using expensive hardware devices (eg. GPUs or TPUs) in order to complete within reasonable amounts of time. To meet this resource demand, dedicated clusters are often provisioned for deep learning~\cite{234916,258957}, with a scheduler that mediates resource sharing between many competing DL jobs. 


Existing schedulers require users to manually configure their jobs, 
which if done improperly, can greatly degrade training performance and resource efficiency. For example, allocating too many GPUs may result in long queuing times and inefficient resource usage, while allocating too few GPUs may result in long runtimes and unused resources. 
Such decisions are especially difficult to make in a shared-cluster setting, since optimal choices are dynamic and depend on the cluster load while a job is running.

Even though recent \textit{elastic} schedulers can automatically select an appropriate amount of resources for each job, they do so blindly to inter-dependent training-related configurations that are just as important.
For example, the \emph{batch size} and \emph{learning rate} of a DL job influence the amount of computation 
needed to train its model. Their optimal choices vary between different DL tasks and model architectures, and they have strong dependence on the job's allocation of resources.

The amount of resources, batch size, and learning rate are difficult to configure appropriately without expert knowledge about both the cluster hardware performance and DL model architecture.
Due to the inter-dependence between their optimal values, they should be configured jointly with each other. Due to the dynamic nature of shared clusters, their optimal values may change over time. This creates a complex web of considerations a user must make in order to configure their job for efficient execution and resource utilization.

\greg{The above paragraph, and really most of the intro from there to "This paper presents" paragraph, kinda struggles with "does the user pick the GPU count or does the scheduler?".  If the user specifies it, then most of what is discussed is much easier... just tune for that number of GPUs.  But, the cluster efficiency may be very poor, because the user can't know what will be available over the job's lifetime and likely even at submission time.  In part as a result, it focuses a lot of the training parameter effects and little on the dynamic GPU availability and where to assign them aspect.  Leaving this note here for now... if folks want, I can try to wade into it Wed evening/night.}\aurick{re-written according to feedback}

How can a cluster scheduler help to automatically configure user-submitted DL jobs? Fundamentally, a properly-configured DL job strikes a balance between two often opposing desires:
(1) \emph{system throughput}, the number of training examples processed per wall-clock time, and
(2) \emph{statistical efficiency}, the amount of progress made per training example processed.

\begin{figure}
\begin{subfigure}{.23\textwidth}
  \centering
  \includegraphics[width=\textwidth, trim=15 12 10 10, clip]{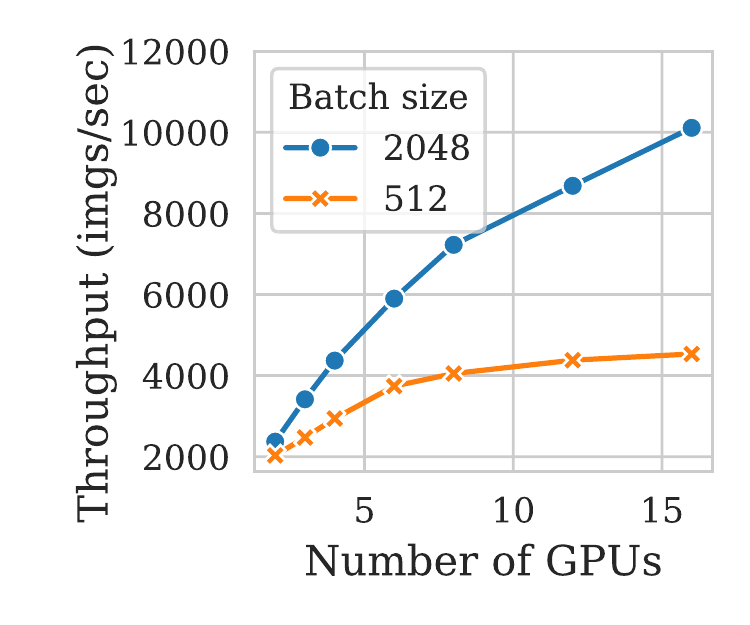}
  \caption{Job scalability (and thus resource utilization) depends on the batch size.}
  \label{fig:intro-scalability}
\end{subfigure}\hspace*{\fill}%
\begin{subfigure}{.23\textwidth}
  \centering
  \includegraphics[width=\textwidth, trim=15 12 10 10, clip]{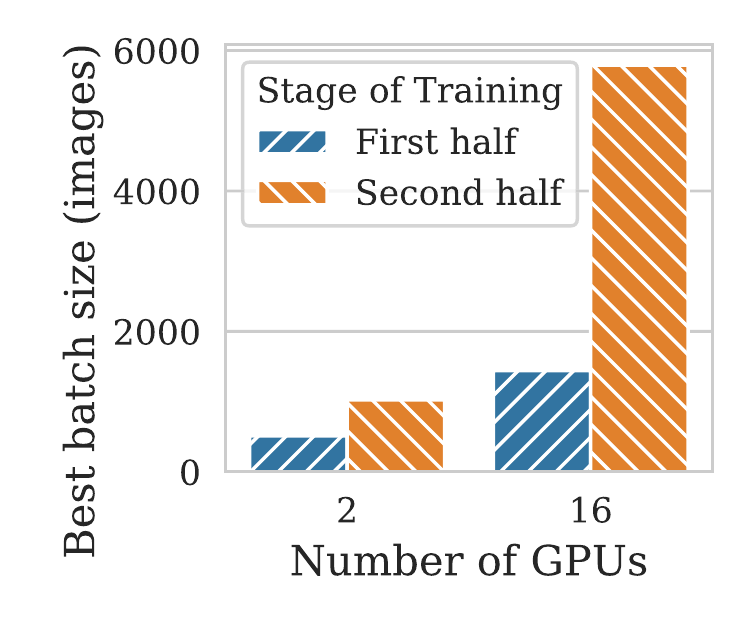}
  \caption{The most efficient batch size depends on the allocated resources and stage of training.}
  \label{fig:intro-batch-size}
\end{subfigure}
\caption{Trade-offs between the batch size, resource scalability, and stage of training (ResNet18 on CIFAR-10). The learning rate is separately tuned for each batch size.}
\vspace{-0.5em}
\label{fig:intro}
\end{figure}

System throughput can be increased by increasing the batch size, as illustrated in Fig.~\ref{fig:intro-scalability}.
A larger batch size enables higher utilization of more compute resources (e.g., more GPUs). 
%
But, even with an optimally-retuned learning rate, increasing the batch size often results in  a decreased statistical efficiency~\cite{DBLP:journals/corr/abs-1811-03600,DBLP:journals/corr/abs-1812-06162}. For every distinct allocation of GPUs, there is potentially a different batch size that best balances increasing system throughput with decreasing statistical efficiency, as illustrated in Fig.~\ref{fig:intro-batch-size}. Furthermore, how quickly the statistical efficiency decreases with respect to the batch size depends on the current training progress. A job in a later stage of training can potentially tolerate 10x or larger batch sizes without degrading statistical efficiency, than earlier during training \cite{DBLP:journals/corr/abs-1812-06162}.

\aurick{Modified the following to bring up more concrete things \oursystem{} tunes.}

Guided by these insights, this paper presents \emph{\oursystem{}}, a hybrid resource scheduler that \emph{co-adaptively} allocates resources and tunes the batch size and learning rate for all DL jobs in a shared cluster. 
\oursystem{} achieves this by jointly managing several system-level and training-related parameters, including the number of GPUs, co-location of workers, per-GPU batch size, gradient accumulation, and learning rate scaling. In particular:

\noindent\(\bigstar\) We propose a formulation of \emph{goodput} for DL jobs, which is a holistic measure of training performance that takes into account both system throughput and statistical efficiency.

\noindent\(\bigstar\) We show that a model of a DL job's goodput can be learned by observing its throughput and statistical behavior during training, and used for predicting the performance given different resource allocations and batch sizes.

\noindent\(\bigstar\) We design and implement a scheduling architecture that uses such models to configure the right combination of resource allocation and training parameters for each pending and running DL job.  This includes locally tuning system-level and training-related parameters for each DL job, and globally optimizing cluster-wide resource allocations. The local and global components actively communicate and cooperate with each other, operating based on the common goal of goodput maximization. 

\noindent\(\bigstar\) We evaluate \oursystem{} on a cluster testbed using a workload derived from a Microsoft cluster trace.
Compared with recent DL schedulers, Tiresias~\cite{227623} and Optimus~\cite{10.1145/3190508.3190517},
\oursystem{} reduces the average job completion time by up to \( 73\% \). Even when all jobs are manually tuned beforehand, \oursystem{} reduces the average job completion time by 37\%--50\%.
At the same time, \oursystem{} improves finish-time fairness~\cite{themis2020} by \( 1.5\times \)--\( 5.4\times \).

\noindent\(\bigstar\) 
We show that, in cloud environments, using goodput-driven auto-scaling based on \oursystem{} can potentially reduce the cost of training large models by 25\%.

\greg{I'm not sure what to do with the cloud thing here.}\aurick{Can maybe remove to save space?}

\section{Background: Distributed DL Training}





Training a deep learning model typically involves minimizing a \emph{loss function} of the form
\begin{equation}
    \mathcal{L}(w) = \frac{1}{|X|} \sum_{x_i \in X} \ell(w, x_i),
\end{equation}
where \( w \in \mathbb{R}^d \) are the model parameters to be optimized, \( X \) is the training dataset, \( x_i \) is an individual sample in \( X \), and \( \ell \) is the loss evaluated at a single sample.

The loss function can be minimized using stochastic gradient descent (SGD) or its variants like AdaGrad~\cite{JMLR:v12:duchi11a} and Adam~\cite{kingma2017adam}. For the purpose of explaining system throughput and statistical efficiency, we will use SGD as the running example. SGD repeatedly applies the following update until the loss converges to a stable value:
    $w^{(t+1)} = w^{(t)} - \eta \hat g^{(t)}$.
\( \eta \) is known as the learning rate, which is a scalar that
controls the magnitude of each update, and 
\( \hat g^{(t)} \) is a stochastic gradient estimate of the loss function
\( \mathcal{L} \), evaluated using a random \emph{mini-batch}
\( \mathcal{M}^{(t)} \subset X\) of the training data:
\begin{equation}
    \hat g^{(t)} = \frac{1}{M} \sum_{x_i \in \mathcal{M}^{(t)}} \nabla \ell(w^{(t)}, x_i).
\end{equation}
The learning rate $\eta$ and batch size \( M=|\mathcal{M}^{(t)}| \) are training parameters which are typically chosen by the user.

\subsection{System Throughput}
The \emph{system throughput} of DL training can be defined as the
number of training samples processed per unit of wall-clock time.
When a DL job is distributed across several nodes, its system throughput
is determined by several factors, including (1)~the allocation and placement
of resources (e.g. GPUs) assigned to the job, (2) the method of distributed execution
and synchronization, and (3) the batch size.

\noindent\textbf{Data-parallel execution.}
\emph{Synchronous data-parallelism} is a popular method of distributed execution for DL training.
The model parameters \( w^{(t)} \) are replicated across a set of distributed GPUs \( 1, \ldots, K \), and each mini-batch \( \mathcal{M}^{(t)} \) is divided into equal-sized partitions per node, \( \mathcal{M}^{(t)}_1, \ldots, \mathcal{M}^{(t)}_K \).
Each GPU \( k \) computes a local gradient estimate \( \hat g^{(t)}_k \) using its own partition:
\begin{equation}
    \hat g^{(t)}_k = \frac{1}{m} \sum_{x_i \in \mathcal{M}^{(t)}_k} \nabla \ell(w^{(t)}, x_i),
\end{equation}
where \( m = |\mathcal{M}^{(t)}_k| \) is the per-GPU batch size. These local gradient estimates are then averaged across all GPUs to obtain the desired \( \hat g^{(t)} \). 
Finally, each node applies the same update using \( \hat g^{(t)} \) to obtain the new model parameters \( w^{(t+1)} \). 

The run-time of each training iteration is determined by two main components.
\emph{First}, the time spent computing each \( \hat g^{(t)}_k \), which we denote by \( T_{grad} \).
\emph{Second}, the time spent averaging \( \hat g^{(t)}_k \) (e.g. using collective all-reduce \cite{DBLP:journals/corr/abs-1802-05799,NIPS2019_9015}) and/or synchronizing \( w^{(t)} \) (e.g. using parameter servers \cite{NIPS2012_4687,DBLP:journals/corr/ChenLLLWWXXZZ15,NIPS2013_4894,216041}) across all GPUs, which we denote by \( T_{sync} \).
\( T_{sync} \)
is influenced by the size of the gradients, performance of the network, and is typically shorter
when the GPUs are co-located within the same physical node or rack.

\noindent\textbf{Limitations due to the batch size.} 
When the number of GPUs is increased, \( T_{grad} \) decreases due to a smaller per-GPU batch size.
On the other hand, \( T_{sync} \), which is typically independent of the batch size, remains unchanged.
By Amdahl's Law, no matter how many GPUs are used, the run-time
of each training iteration is lower bounded by \( T_{sync} \).
To overcome this scalability limitation, a common strategy is to
increase the batch size. Doing so causes the 
local gradient estimates to be computed over more training examples and thereby increasing the ratio of \( T_{grad} \) to \( T_{sync} \).
As a result, using a larger batch size enables higher system throughput
when scaling to more GPUs in the synchronous data-parallel setting.

\subsection{Statistical Efficiency}
\label{sec:statistical-efficiency}


The \emph{statistical efficiency} of DL training can be defined as the amount of
training progress made per unit of training data processed, influenced by parameters such as \emph{batch size} or \emph{learning rate}; for example, a larger batch size normally decreases the statistical efficiency.
The ability to predict statistical efficiency is key to
improving said statistical efficiency, because we can use the predictions to better adapt the batch sizes and learning rates.

\noindent\textbf{Gradient noise scale.}
Previous work \cite{DBLP:journals/corr/abs-1812-06162, johnson2020adascale} relate the statistical efficiency
of DL training to the \textit{gradient noise scale} (GNS), which measures the noise-to-signal ratio of the stochastic gradient.
%
%
%
A larger GNS means that training parameters such as the
batch size and learning rate can be increased to higher values
with relatively less reduction of the statistical efficiency.
%
The GNS can vary greatly between different
DL models \cite{DBLP:journals/corr/abs-1811-12941}.
It is also non-constant and tends to gradually increase during training,
by up to \( 10 \times \) or more \cite{DBLP:journals/corr/abs-1812-06162}.
Thus, it is possible to attain significantly better statistical efficiency for large batch sizes
later on during training.

The gradient noise scale mathematically captures an intuitive explanation of how the batch size affects statistical efficiency. When the stochastic gradient has low noise, adding more training examples to each mini-batch does not significantly improve each gradient estimate, which lowers statistical efficiency. When the  stochastic gradient has high noise, adding more training examples to each mini-batch reduces the noise of each gradient estimate, which maintains high statistical efficiency. Near convergence, the stochastic gradients have relatively lower signal than noise, and so larger batch sizes can be more useful later in training.



\noindent\textbf{Learning rate scaling.}
When training with an increased total batch size $M$, the learning rate $\eta$ should also be increased, otherwise the final trained model quality/accuracy can be significantly worse~\cite{DBLP:journals/corr/abs-1811-03600}.
How to increase the learning rate varies between different models and training algorithms (e.g. SGD, Adam~\cite{kingma2017adam}, AdamW~\cite{loshchilov2019decoupled}), and several well-established scaling rules may be used. 
For example, the linear scaling rule \cite{DBLP:journals/corr/GoyalDGNWKTJH17}, which prescribes that \( \eta \) be scaled proportionally with \( M \), or the square-root scaling rule~\cite{DBLP:journals/corr/Krizhevsky14, DBLP:journals/corr/abs-1901-08256} (commonly used with Adam), which prescribes that \( \eta \) be scaled proportionally with \( \sqrt M \).
More recent scaling rules such as AdaScale~\cite{johnson2020adascale} may scale the learning rate adaptively during training.

In addition to decreasing statistical efficiency, using large batch sizes may also degrade the final model quality in terms of validation performance~\cite{DBLP:journals/corr/KeskarMNST16,Smith2018ABP,DBLP:journals/corr/abs-1811-12941}, although the reasons behind this effect are not completely understood at the time of this paper. However, for each of the learning rate scaling rules mentioned above, there is usually a problem-dependent range of batch sizes that achieve similar validation performances. Within these ranges, the batch size may be chosen more freely without significantly degrading the final model quality.

\subsection{Existing DL Schedulers}
\label{sec:existing-schedulers}



We broadly group existing DL schedulers into two categories, to put \oursystem{} in context. 
First, \emph{non-scale-adaptive} schedulers are agnostic to the performance scalability of DL jobs with respect to the amount of allocated resources. 
For example, Tiresias~\cite{227623} requires users to specify the number of GPUs at the time of job submission, which will be fixed for the lifetime of the job. 
Gandiva~\cite{222611} 
also 
requires users to specify number of GPUs, but enhances resource utilization through fine-grained time sharing and job packing. 
Although Gandiva may dynamically change the number of GPUs used by a job, it does so opportunistically and not based on knowledge of job scalability.

Second, \emph{scale-adaptive} schedulers automatically decide the amount of resources allocated to each job based on how well they can be utilized to speed up the job. 
For example, Optimus~\cite{10.1145/3190508.3190517} learns a predictive model for the system throughput of each job given various amounts of resources, and optimizes cluster-wide resource allocations to minimize the average job completion time. 
SLAQ~\cite{10.1145/3127479.3127490}, which was not evaluated on DL, uses a similar technique to minimize the average loss values for training general ML models.
Gavel~\cite{258896} goes further by scheduling based on a throughput metric that is comparable across different accelerator types.\footnote{\oursystem{}'s current throughput model does not consider accelerator heterogeneity.  We believe that extending with Gavel's metric would allow \oursystem{} to co-adapt for goodput in heterogeneous DL clusters.}
AntMan~\cite{258957} uses dynamic scaling and fine-grained GPU sharing to improve cluster utilization, resource fairness, and job completion times.
Themis~\cite{themis2020} introduces the notion of finish-time fairness, and promotes fairness between multiple DL applications with a two-level scheduling architecture.

Crucially, existing schedulers are agnostic to the statistical efficiency of DL training and the inter-dependence of resource decisions and training parameters.
\oursystem{} explicitly co-adapts these inter-dependent values to improve goodput for DL jobs.
%


\section{The Goodput of DL Training and \oursystem{}}
\label{sec:goodput}

In this section, we define the \emph{goodput}\footnote{Our notion of goodput for DL is analogous to the traditional definition of goodput in computer networks, ie. the \emph{useful} portion of throughput as benchmarked by training progress per unit of wall-clock time.} of DL jobs, which is a measure of training performance that takes into account both system throughput and statistical efficiency. We then describe how the goodput can be measured during training and used as a predictive model, which is leveraged by \oursystem{} to jointly optimize cluster-wide resource allocations and batch sizes.


\begin{definition}{(Goodput)}
The \emph{goodput} of a DL training job at iteration \( t \) is the product between its system throughput and its statistical efficiency at iteration \( t \),
\begin{equation}
    \small
    \mathtt{GOODPUT}_t(\star) = \mathtt{THROUGHPUT}(\star) \times \mathtt{EFFICIENCY}_t\left(M(\star)\right),
    \label{eqn:goodput}
\end{equation}
where \( \star \) represents any configuration parameters that jointly influence the throughput and batch size during training, and \( M \) is the total batch size summed across all allocated GPUs.
\end{definition}

While the above definition is general across many training systems, we focus on three configuration parameters of particular impact in the context of efficient resource scheduling, i.e. $ \star = (a, m, s) $, where:

\begin{itemize}
    \item \( a \in \mathbb{Z}^N \): the \emph{allocation vector}, where \( a_n \) is the number of GPUs allocated from node \( n \).
    \item \( m \in \mathbb{Z} \): the \textit{per-GPU batch size}.
    \item \( s \in \mathbb{Z} \): number of \textit{gradient accumulation steps} (\S\ref{sec:modeling-system-throughput}).
\end{itemize}
The total batch size is then defined as
\[ M(a, m, s) = \mathtt{SUM}(a) \times m \times (s + 1). \]

\noindent\textbf{\oursystem's approach.}
An initial batch size \( M_0 \) and learning rate (LR) \( \eta_0 \) are selected by the user when submitting their job. \oursystem{} will start each job using a single GPU, \( m = M = M_0 \), \( s = 0 \), and \( \eta = \eta_0 \).
As the job runs, \oursystem{} profiles its execution to learn and refine predictive models for both \( \mathtt{THROUGHPUT} \) (\S\ref{sec:modeling-system-throughput}) and \( \mathtt{EFFICIENCY} \) (\S\ref{sec:modeling-statistical-efficiency}).
Using these predictive models, \oursystem{} periodically re-tunes $(a, m, s)$ for each job, according to cluster-wide resource availability and performance (\S\ref{sec:sched}).

\( \mathtt{EFFICIENCY}_t \) is measured \emph{relative to} the initial batch size \( M_0 \) and learning rate \( \eta_0 \), and \oursystem{} only considers batch sizes that are at least the initial batch size, ie. \( M \ge M_0 \). In this scenario, \( \mathtt{EFFICIENCY}_t(M) \) is a fraction (between \( 0 \) and \( 1 \)) relative to \( \mathtt{EFFICIENCY}_t(M_0) \). Therefore, goodput can be interpreted as the portion of the throughput that is useful for training progress, being equal to the throughput if and only if perfect statistical efficiency is achieved.

\begin{figure*}
\begin{subfigure}{.21\textwidth}
  \centering
  \includegraphics[width=\textwidth,clip,trim=-15 8 8 8] {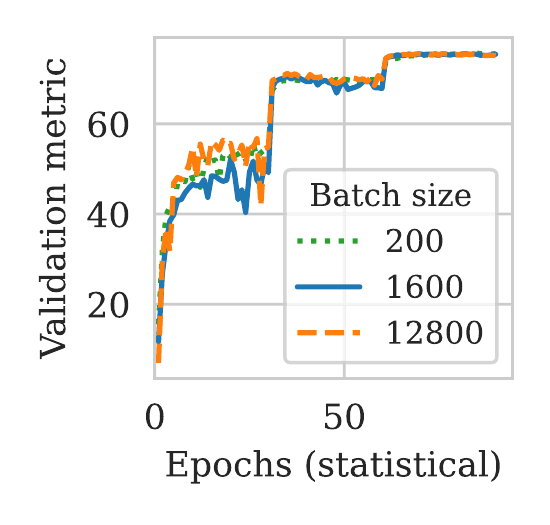}
\end{subfigure}\hspace*{\fill}%
\begin{subfigure}{.155\textwidth}
  \centering
  \includegraphics[width=\textwidth,clip,trim=25 10 10 10] {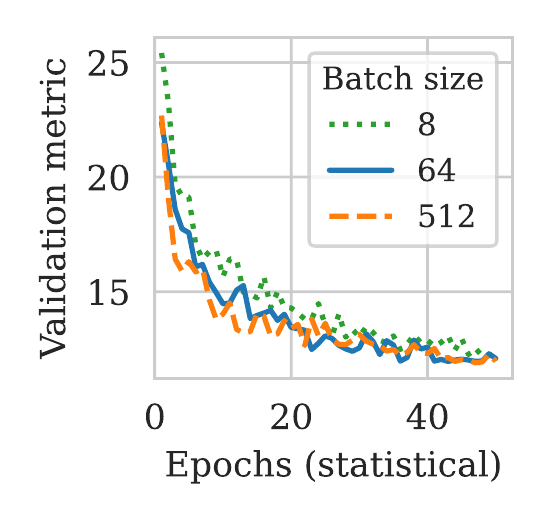}
\end{subfigure}\hspace*{\fill}%
\begin{subfigure}{.155\textwidth}
  \centering
  \includegraphics[width=\textwidth,clip,trim=25 10 10 10] {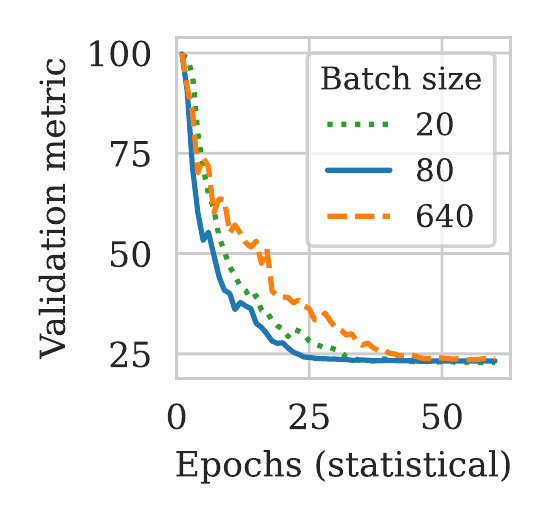}
\end{subfigure}\hspace*{\fill}%
\begin{subfigure}{.155\textwidth}
  \centering
  \includegraphics[width=\textwidth,clip,trim=25 10 10 10] {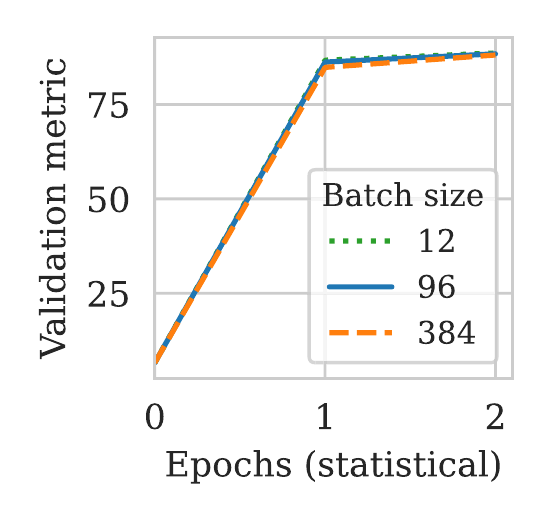}
\end{subfigure}\hspace*{\fill}%
\begin{subfigure}{.155\textwidth}
  \centering
  \includegraphics[width=\textwidth,clip,trim=25 10 10 10] {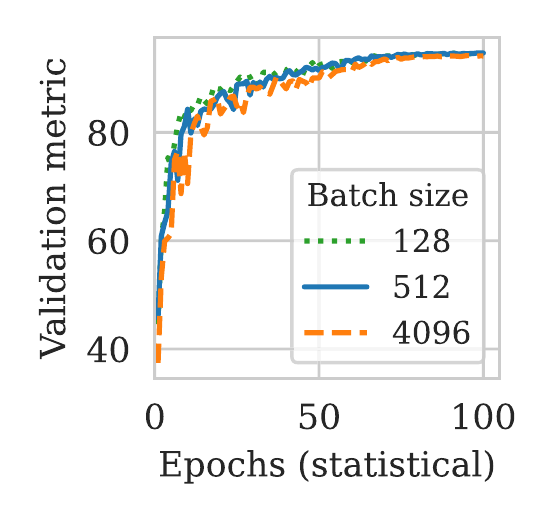}
\end{subfigure}\hspace*{\fill}%
\begin{subfigure}{.155\textwidth}
  \centering
  \includegraphics[width=\textwidth,clip,trim=25 10 10 10] {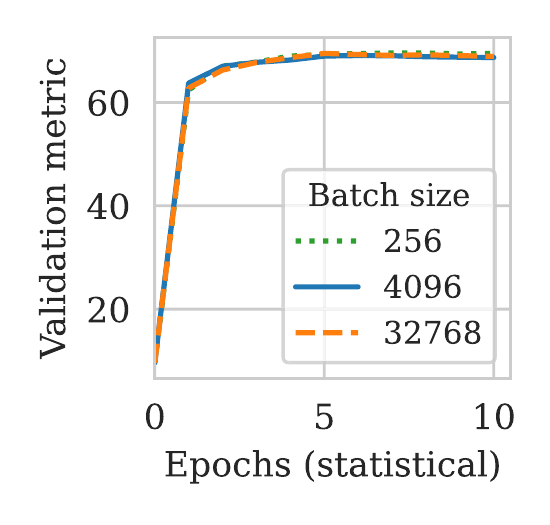}
\end{subfigure}
\begin{subfigure}{.212\textwidth}
  \centering
  \includegraphics[width=\textwidth,clip,trim=8 8 8 8] {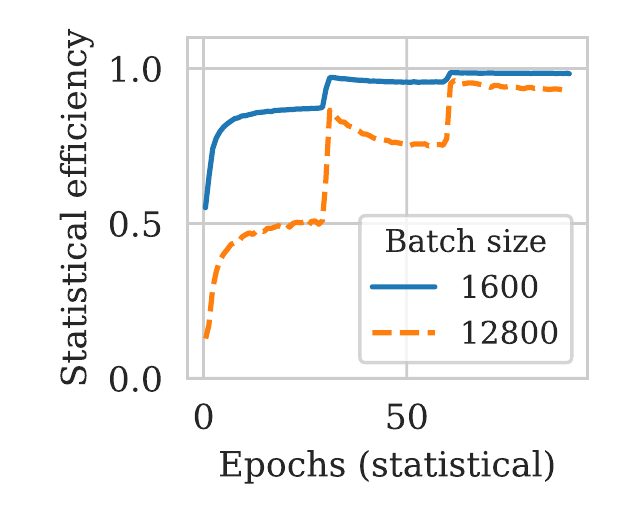}
\end{subfigure}\hspace*{\fill}%
\begin{subfigure}{.155\textwidth}
  \centering
  \includegraphics[width=\textwidth,clip,trim=50 10 10 10] {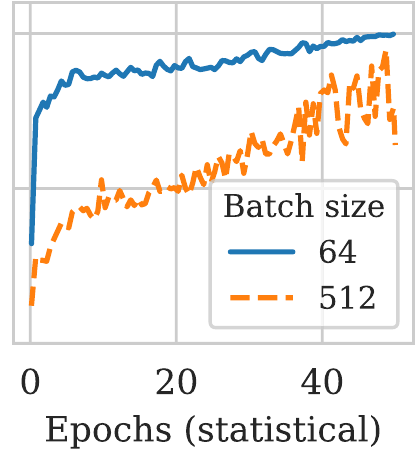}
\end{subfigure}\hspace*{\fill}%
\begin{subfigure}{.155\textwidth}
  \centering
  \includegraphics[width=\textwidth,clip,trim=50 10 10 10] {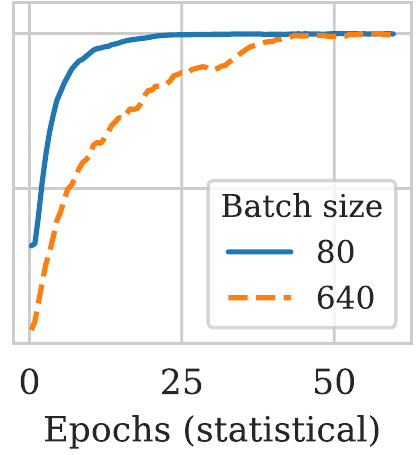}
\end{subfigure}\hspace*{\fill}%
\begin{subfigure}{.155\textwidth}
  \centering
  \includegraphics[width=\textwidth,clip,trim=50 10 10 10] {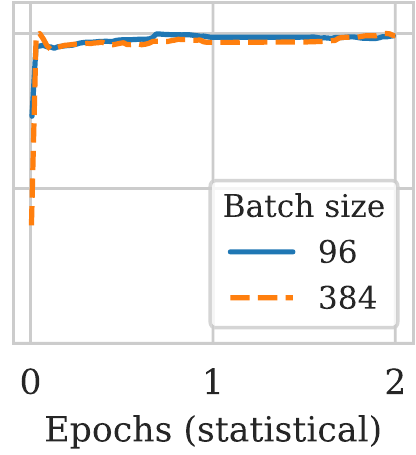}
\end{subfigure}\hspace*{\fill}%
\begin{subfigure}{.155\textwidth}
  \centering
  \includegraphics[width=\textwidth,clip,trim=50 10 10 10] {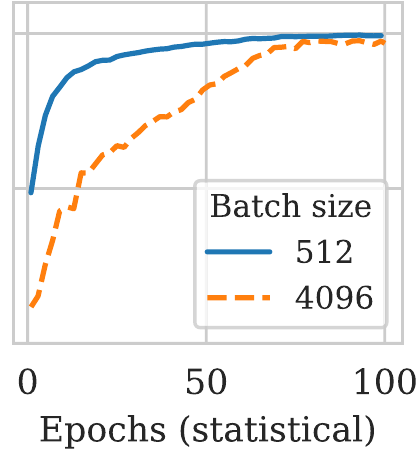}
\end{subfigure}\hspace*{\fill}%
\begin{subfigure}{.155\textwidth}
  \centering
  \includegraphics[width=\textwidth,clip,trim=50 10 10 10] {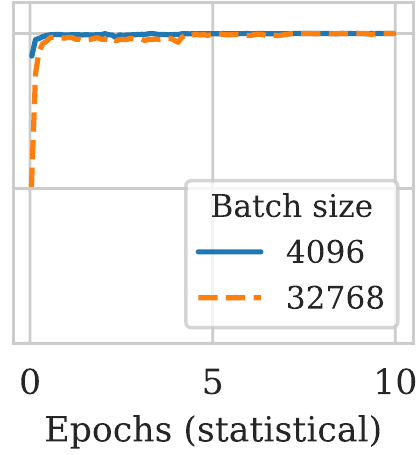}
\end{subfigure}
\begin{subfigure}{.212\textwidth}
  \centering
  \includegraphics[width=\textwidth,clip,trim=8 10 8 8] {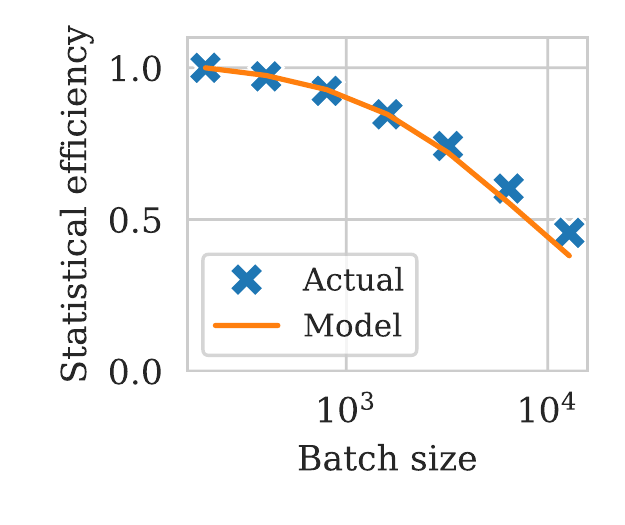}
  \caption{ImageNet}
  \label{fig:efficiency-vs-bsz-imagenet}
\end{subfigure}\hspace*{\fill}%
\begin{subfigure}{.155\textwidth}
  \centering
  \includegraphics[width=\textwidth,clip,trim=50 12 10 10] {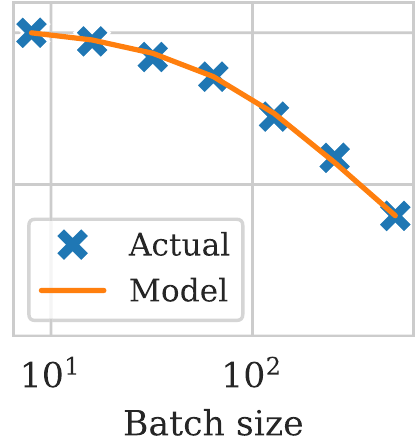}
  \caption{YoloV3}
  \label{fig:efficiency-vs-bsz-yolov3}
\end{subfigure}\hspace*{\fill}%
\begin{subfigure}{.155\textwidth}
  \centering
  \includegraphics[width=\textwidth,clip,trim=50 12 10 10] {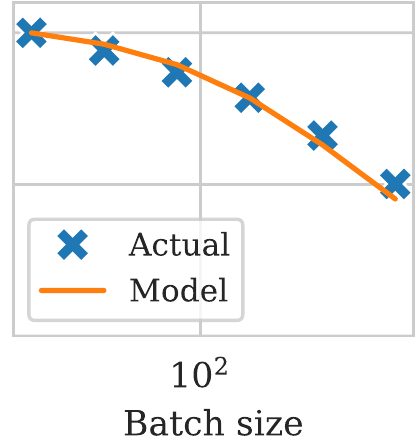}
  \caption{DeepSpeech2}
  \label{fig:efficiency-vs-bsz-deepspeech2}
\end{subfigure}\hspace*{\fill}%
\begin{subfigure}{.155\textwidth}
  \centering
  \includegraphics[width=\textwidth,clip,trim=50 12 10 10] {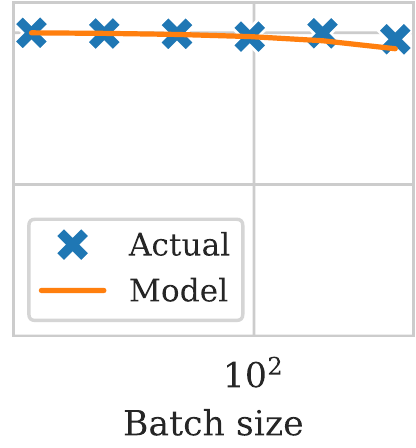}
  \caption{BERT (fine-tune)}
  \label{fig:efficiency-vs-bsz-bert}
\end{subfigure}\hspace*{\fill}%
\begin{subfigure}{.155\textwidth}
  \centering
  \includegraphics[width=\textwidth,clip,trim=50 12 10 10] {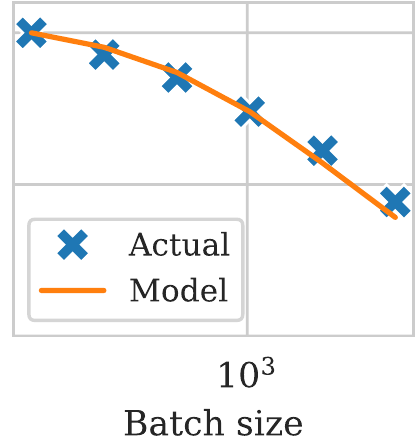}
  \caption{CIFAR10}
  \label{fig:efficiency-vs-bsz-cifar10}
\end{subfigure}\hspace*{\fill}%
\begin{subfigure}{.155\textwidth}
  \centering
  \includegraphics[width=\textwidth,clip,trim=50 12 10 10] {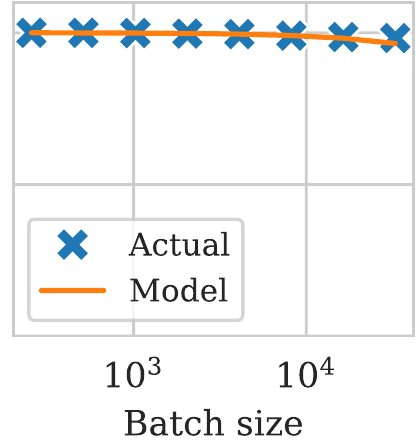}
  \caption{Recommendation}
  \label{fig:efficiency-vs-bsz-ncf}
\end{subfigure}
\caption{Statistical efficiency for all models described in Table~\ref{tab:models}. TOP: validation metric vs training progress for three different batch sizes: \( M_0 \), an intermediate batch size, and the max batch size limit we set for each DL task. Metrics are as defined in Table~\ref{tab:models} except for YoloV3 for which validation loss is shown. MIDDLE: measured statistical efficiency vs. training progress for two different batch sizes. Training progress (x-axis) in the top two rows is shown in terms of ``statistical epochs'', defined as \( \frac{M}{|X|} \sum_t \mathtt{EFFICIENCY}_t(M) \) where \( |X| \) is the size of the training dataset. BOTTOM: measured \( \mathtt{EFFICIENCY}_t \) vs. predicted \( \mathtt{EFFICIENCY}_t \) for a range of batch sizes (log-scaled), using \( \varphi_t \) measured using the median batch size from each range, during an early-training epoch (roughly 1/8th of the way through training).
}
\label{fig:efficiency-model}
\end{figure*}

\noindent\textbf{Plug-in Learning Rate Scaling.}
Recall from \S\ref{sec:statistical-efficiency} that different training jobs may require different learning rate scaling rules to adjust $\eta$ in response to changes in \( M \). In order to support a wide variety of LR scaling rules, including state-of-the-art rules such as AdaScale~\cite{johnson2020adascale}, \oursystem{} provides a plug-in interface that can be implemented using a function signature
\[ \mathtt{SCALE\_LR}(M_0, M) \longrightarrow \lambda. \]
\( \mathtt{SCALE\_LR} \) is called before every model update step,
and \( \lambda \) is used by \oursystem{} to scale the learning rate. The implementation of \( \mathtt{SCALE\_LR} \) can utilize metrics collected during training, such as the gradient noise scale. Using this interface, one can implement rules including AdaScale, square-root scaling~\cite{DBLP:journals/corr/Krizhevsky14},
linear scaling~\cite{DBLP:journals/corr/GoyalDGNWKTJH17} and LEGW~\cite{DBLP:journals/corr/abs-1901-08256}.

\subsection{Modeling Statistical Efficiency}
\label{sec:modeling-statistical-efficiency}

We model \( \mathtt{EFFICIENCY}_t(M) \) as the amount of progress made per training example using $M$, relative to using $M_0$. For SGD-based training, this quantity can be expressed in terms of the gradient noise scale (GNS)~\cite{DBLP:journals/corr/abs-1812-06162}.
To support popular adaptive variants of SGD like Adam~\cite{kingma2017adam} and AdaGrad~\cite{ward2019adagrad}, we use the \textit{pre-conditioned gradient noise scale} (PGNS), derived by closely following the original derivation of the GNS (``simple'' noise scale in \cite{DBLP:journals/corr/abs-1812-06162}) starting from pre-conditioned SGD\footnote{Pre-conditioned SGD optimizes \( \mathcal{L}(Pw) \) instead of \( \mathcal{L}(w) \), where \( P \) is known as a pre-conditioning matrix. Adaptive variants of SGD such as Adam and AdaGrad may be viewed as vanilla SGD (with momentum) applied together with a particular pre-conditioning matrix \( P \).} rather than vanilla SGD. The PGNS, which we denote by \( \varphi_t \), is expressed as
\begin{equation}
    \varphi_t = \frac{\mathrm{tr}(P\Sigma P^T)}{|Pg|^2},
\end{equation}
where \( g \) is the true gradient, \( P \) is the pre-conditioning matrix of the adaptive SGD algorithm, and \( \Sigma \) is the covariance matrix of per-example stochastic gradients. The PGNS is a generalization of the GNS and is mathematically equivalent to the GNS for the special case of vanilla SGD.

Similar to the GNS (Appendix D of \cite{DBLP:journals/corr/abs-1812-06162}), it takes \( 1 + \varphi_t/M \) training iterations to make a similar amount of training progress across different batch sizes \( M \). Therefore, we can use the PGNS \( \varphi_t \) to define a concrete expression for \( \mathtt{EFFICIENCY}_t(M) \) as
\begin{equation}
    \mathtt{EFFICIENCY}_t(M) = \frac{\varphi_t + M_0}{\varphi_t + M}.
    \label{eqn:statistical-efficiency}
\end{equation}
%
%

Intuitively, Eqn.~\ref{eqn:statistical-efficiency} measures the contribution from each training example to the overall progress. If \( \mathtt{EFFICIENCY}_t(M) = E \), then (1) \( 0 < E \le 1 \), and (2) training using batch size \( M \) will need to process \( 1/E \) times as many training examples to make the same progress as using batch size \( M_0 \).

During training, \oursystem{} estimates the value of
\( \varphi_t \),
then uses Eqn~\ref{eqn:statistical-efficiency} to predict the
\( \mathtt{EFFICIENCY}_t \) at different batch sizes. The measured
value of \( \varphi_t \) varies
according to the training progress at iteration \( t \), thus
\( \mathtt{EFFICIENCY}_t(M) \) reflects the lifetime-dependent trends
exhibited by the true statistical efficiency.

Fig.~\ref{fig:efficiency-model} (TOP) shows the validation metrics on a held-out dataset for a variety of DL training tasks (details in Table ~\ref{tab:models}) versus their training progress. ``Statistical epochs''\footnote{Similar to the notion of ``scale-invariant iterations'' defined in \cite{johnson2020adascale}.} is the number of training iterations normalized by \( \mathtt{EFFICIENCY}_t \) so that each statistical epoch makes theoretically, as projected by our model, the same training progress across different batch sizes. Thus, the degree of similarity between validation curves at different batch sizes is an indicator for the accuracy of \( \mathtt{EFFICIENCY}_t \) as a predictor of actual training progress.

Although there are differences in the validation curves for several DL tasks (especially in earlier epochs), they achieve similar best values across the different batch sizes we evaluated (\( \pm 1 \)\% relative difference for all tasks except DeepSpeech2 at \( \pm 4 \)\%). We note that these margins are within the plateau of high-quality models expected from large-batch training~\cite{masters2018revisiting}.

Fig.~\ref{fig:efficiency-model} (MIDDLE and BOTTOM) show the measured and predicted \( \mathtt{EFFICIENCY}_t \) during training and for a range of different batch sizes. In general, larger batch sizes have lower \( \mathtt{EFFICIENCY}_t \) early in training, but close the gap later on in training. The exceptions being BERT, which is a fine-tuning task starting from an already pre-trained model, and recommendation, which uses a much smaller and shallower model architecture than the others. How \( \mathtt{EFFICIENCY}_t \) changes during training varies from task to task, and depends on specific properties like the learning rate schedule. For example, \( \mathtt{EFFICIENCY}_t \) for ImageNet, which uses step-based learning rate annealing, experiences sharp increases whenever the learning rate is annealed.

Finally, we note that the \( \mathtt{EFFICIENCY}_t \) function (which is supplied with estimates of $\varphi_t$ by \oursystem{}) is able to accurately model observed values at a range of different batch sizes.
This means that \( \varphi_t \) measured using batch size \( M \) can be used by \oursystem{} to predict the value of \( \mathtt{EFFICIENCY}_t \) at a different batch size \( M^\prime \) without needing to train using \( M^\prime \) ahead of time.

\noindent\textbf{Upper batch size limit.} In some cases, as the batch size increases, the chosen LR scaling rule may break down before the statistical efficiency decreases, which degrades the final model quality. To address these cases, the application may define a maximum batch size limit that will be respected by \oursystem{}. Nevertheless, we find that a batch size up to \(32\times\) larger works well in most cases. Furthermore, limits for common models are well-studied for popular LR scaling rules~\cite{johnson2020adascale,DBLP:journals/corr/GoyalDGNWKTJH17,DBLP:journals/corr/abs-1901-08256,DBLP:journals/corr/abs-1811-03600}. As better LR scaling rules are developed, they may be incorporated into \oursystem{} using its plug-in interface (\S\ref{sec:goodput}).



\noindent\textbf{Estimating \( \varphi_t \).} The PGNS \( \varphi_t \) can be estimated in a similar fashion as the GNS by following Appendix A.1 of \cite{DBLP:journals/corr/abs-1812-06162}, except using the pre-conditioned gradient \( Pg \) instead of the gradient \( g \). This can be done efficiently when there are multiple data-parallel processes by using the different values of \( \hat g_k^{(t)} \) already available on each GPU \( k \). 
However, this method doesn't work when there is only a single GPU (and gradient accumulation is off, i.e. \( s=0 \)). In this particular situation, \oursystem{} switches to a differenced variance estimator \cite{WANG2017125} which uses consecutive gradient estimates \( \hat g^{(t-1)} \) and \( \hat g^{(t)} \).

\begin{figure*}
\begin{subfigure}{.184\textwidth}
  \centering
  \includegraphics[width=\textwidth,clip,trim=8 8 8 8] {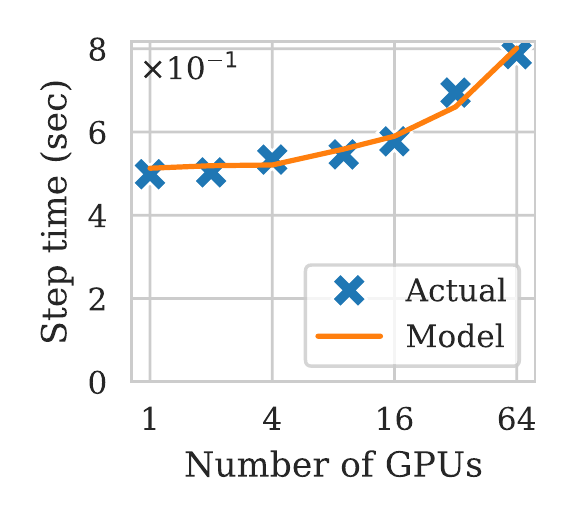}
\end{subfigure}\hspace*{\fill}%
\begin{subfigure}{.163\textwidth}
  \centering
  \includegraphics[width=\textwidth,clip,trim=22 10 10 10] {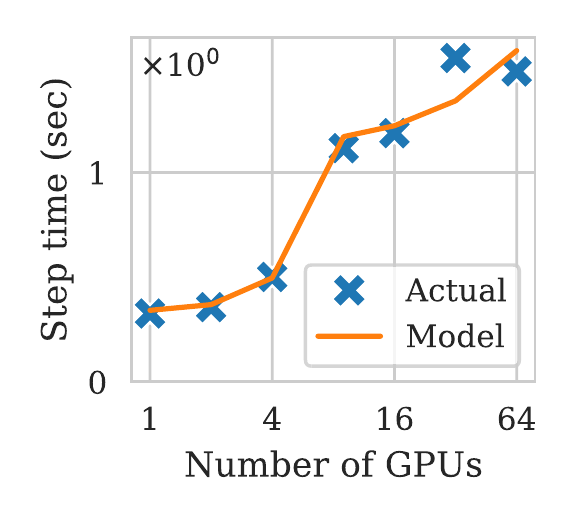}
\end{subfigure}\hspace*{\fill}%
\begin{subfigure}{.163\textwidth}
  \centering
  \includegraphics[width=\textwidth,clip,trim=22 10 10 10] {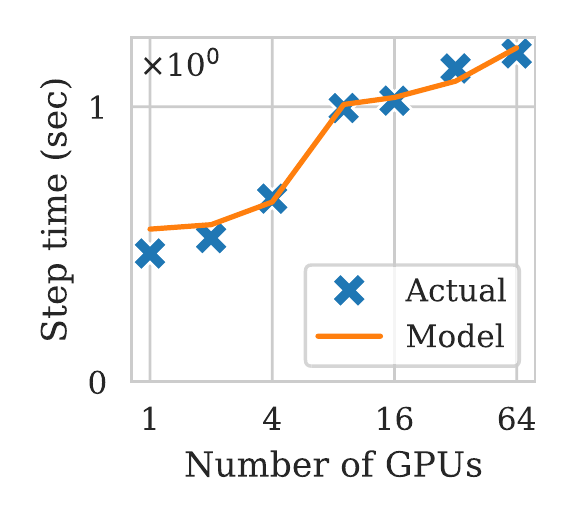}
\end{subfigure}\hspace*{\fill}%
\begin{subfigure}{.163\textwidth}
  \centering
  \includegraphics[width=\textwidth,clip,trim=22 10 10 10] {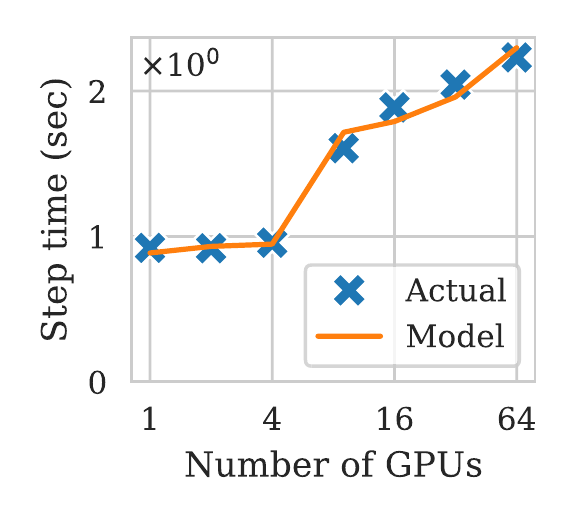}
\end{subfigure}\hspace*{\fill}%
\begin{subfigure}{.163\textwidth}
  \centering
  \includegraphics[width=\textwidth,clip,trim=22 10 10 10] {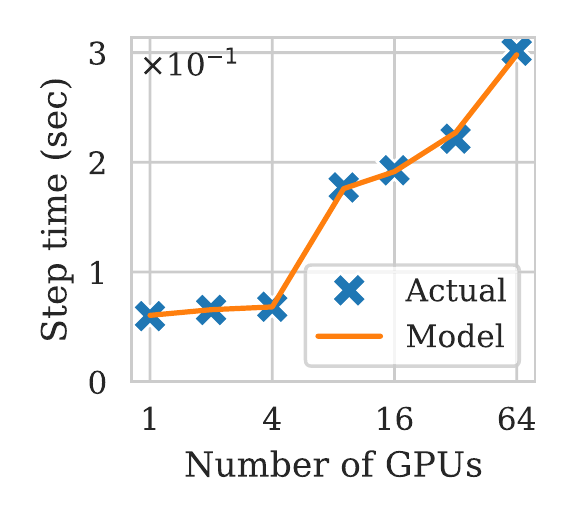}
\end{subfigure}\hspace*{\fill}%
\begin{subfigure}{.163\textwidth}
  \centering
  \includegraphics[width=\textwidth,clip,trim=22 10 10 10] {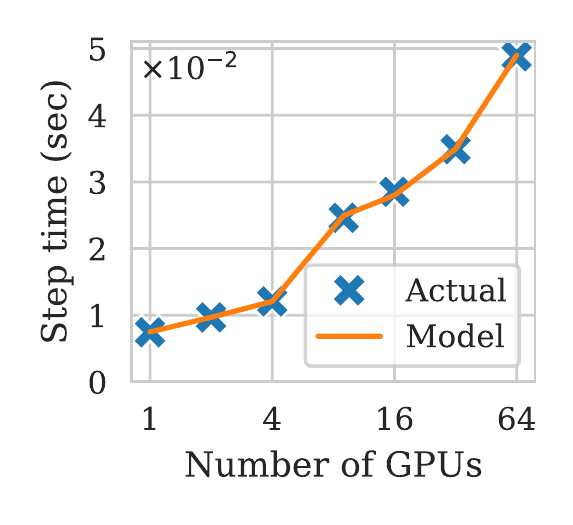}
\end{subfigure}
\begin{subfigure}{.184\textwidth}
  \centering
  \includegraphics[width=\textwidth,clip,trim=8 10 8 8] {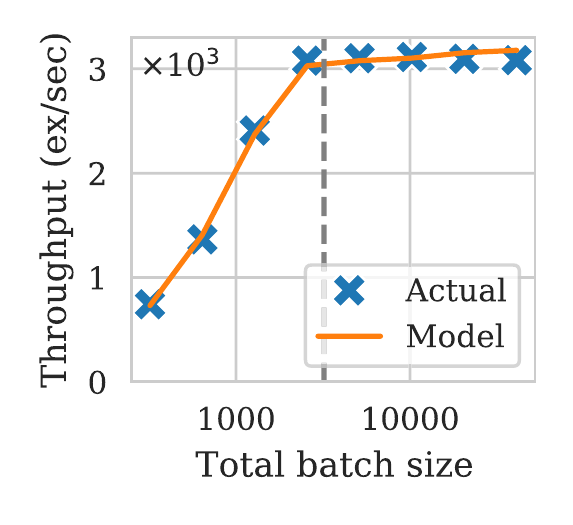}
  \caption{ImageNet}
  \label{fig:throughput-vs-bsz-imagenet}
\end{subfigure}\hspace*{\fill}%
\begin{subfigure}{.163\textwidth}
  \centering
  \includegraphics[width=\textwidth,clip,trim=22 12 10 10] {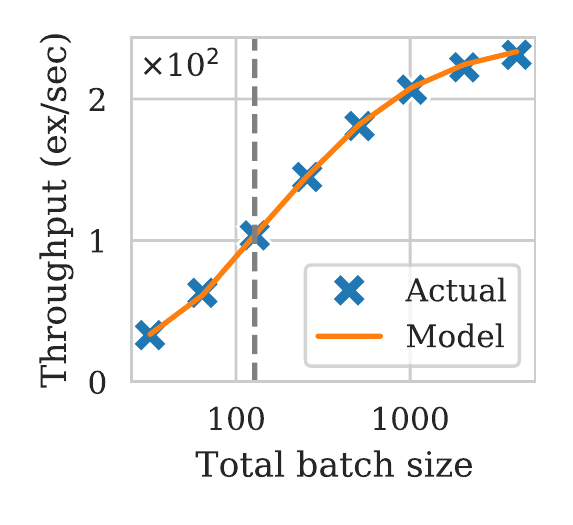}
  \caption{YoloV3}
  \label{fig:throughput-vs-bsz-yolov3}
\end{subfigure}\hspace*{\fill}%
\begin{subfigure}{.163\textwidth}
  \centering
  \includegraphics[width=\textwidth,clip,trim=22 12 10 10] {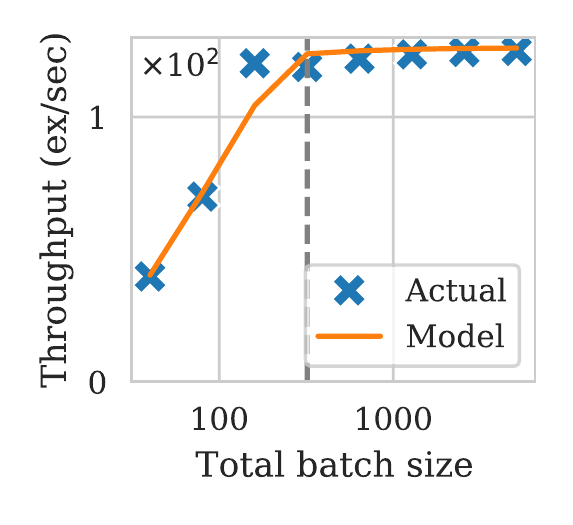}
  \caption{DeepSpeech2}
  \label{fig:throughput-vs-bsz-deepspeech2}
\end{subfigure}\hspace*{\fill}%
\begin{subfigure}{.163\textwidth}
  \centering
  \includegraphics[width=\textwidth,clip,trim=22 12 10 10] {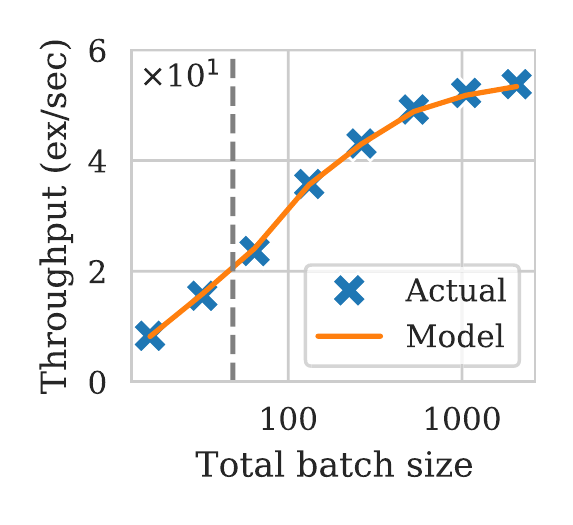}
  \caption{BERT (fine-tune)}
  \label{fig:throughput-vs-bsz-bert}
\end{subfigure}\hspace*{\fill}%
\begin{subfigure}{.163\textwidth}
  \centering
  \includegraphics[width=\textwidth,clip,trim=22 12 10 10] {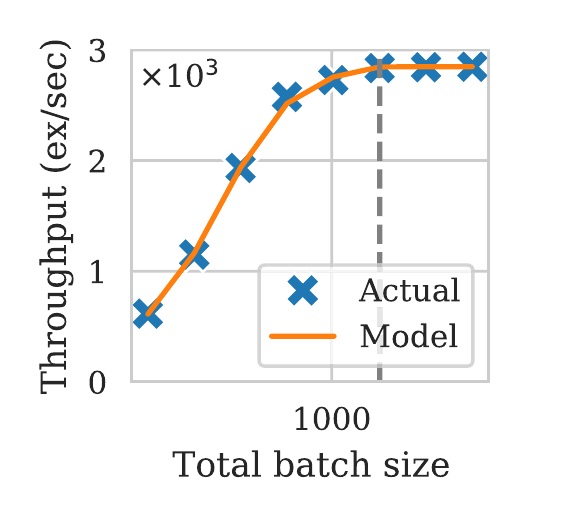}
  \caption{CIFAR10}
  \label{fig:throughput-vs-bsz-cifar10}
\end{subfigure}\hspace*{\fill}%
\begin{subfigure}{.163\textwidth}
  \centering
  \includegraphics[width=\textwidth,clip,trim=22 12 10 10] {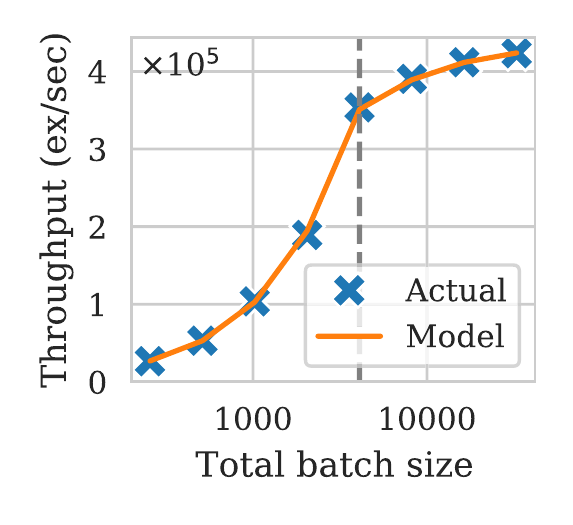}
  \caption{Recommendation}
  \label{fig:throughput-vs-bsz-ncf}
\end{subfigure}
\caption{System throughput for all models described in Table~\ref{tab:models}, as measured using g4dn.12xlarge instances in AWS each with 4 NVIDIA T4 GPUs and created within the same placement group. Eqn.~\ref{eqn:accumulation} was fitted using the observed data that appeared in each plot. TOP: time per training iteration vs. the number of allocated GPUs (log-scaled), with the per-GPU batch size held constant. The GPUs are placed in as few 4-GPU nodes as possible, which causes a sharp increase beyond 4 GPUs (when inter-node network synchronization becomes required). BOTTOM: system throughput (examples per second) vs. total batch size (log-scaled), with the number of GPUs held constant. To the left of the vertical dashed line, the entire mini-batch fits within GPU memory. To the right, the total batch size is achieved using gradient accumulation.}
\label{fig:throughput-model}
\end{figure*}

\subsection{Modeling System Throughput}
\label{sec:modeling-system-throughput}


To model and predict the system throughput for data-parallel DL, we aim to predict the time spent per training iteration, \( T_{iter} \), and then calculate the throughput as
\begin{equation}
    \mathtt{THROUGHPUT}(a, m, s) = M(a, m, s) / T_{iter}(a, m, s).
    \label{eqn:throughput-model}
\end{equation}
We start by separately modeling \( T_{grad} \), the time in each iteration spent computing local gradient estimates, and \( T_{sync} \), the time in each iteration spent averaging gradient estimates and synchronizing model parameters across all GPUs. We also start by assuming no gradient accumulation, i.e. \( s = 0 \).

\noindent\textbf{Modeling \( T_{grad} \).} 
The local gradient estimates are computed using back-propagation, whose run-time scales linearly with the per-GPU batch size \( m \). Thus, we model \( T_{grad} \) as
\begin{equation}
    T_{grad}(m) = \alpha_{grad} + \beta_{grad} \cdot m,
\end{equation}
where \( \alpha_{grad}, \beta_{grad} \) are fittable parameters.

\noindent\textbf{Modeling \( T_{sync} \).}
When allocated a single GPU, no synchronization is needed and \( T_{sync} = 0 \).
Otherwise, we model \( T_{sync} \) as a linear function of the number of GPUs since in data-parallelism, the amount of data sent and received from each replica is typically only dependent on the size of the gradients and/or parameters. We include a linear factor to account for performance retrogressions associated with using three or more GPUs, such as increasing likelihood of stragglers or network delays.

Co-location of GPUs on the same node reduces network communication, which can improve \( T_{sync} \). Thus, we use different parameters depending on GPU placement.
Letting \( K = \mathtt{SUM}(a) \) be the number of allocated GPUs,
\begin{equation}
    T_{sync}(a, m) =
    \begin{cases}
        0 &\text{if } K = 1 \\
        \alpha_{sync}^{local} + \beta_{sync}^{local} \cdot (K-2) &\text{if } N = 1\text{, } K \ge 2 \\
        \alpha_{sync}^{node} + \beta_{sync}^{node} \cdot (K-2) & \text{otherwise,}
    \end{cases}
\end{equation}
where \( N \) is the number of physical nodes occupied by at least one replica. \( \alpha_{sync}^{local} \) and \(  \beta_{sync}^{local} \) are the constant and retrogression parameters for when all processes are co-located onto the same node. \( \alpha_{sync}^{node} \) and \(  \beta_{sync}^{node} \) are the analogous parameters for when at least two process are located on different nodes. Note that our model for \( T_{sync} \) can be extended to account for rack-level locality by adding a third pair of parameters.

\noindent\textbf{Combining \( T_{grad} \) and \( T_{sync} \).} Modern DL frameworks can partially overlap \( T_{grad} \) and \( T_{sync} \) by overlapping gradient computation with network communication~\cite{203269}. The degree of this overlap depends on structures in the specific DL model being trained, like the ordering and sizes of its layers.

Assuming no overlap, then \( T_{iter} = T_{grad} + T_{sync} \). Assuming perfect overlap, then \( T_{iter} = \max(T_{grad}, T_{sync}) \). A realistic value of \( T_{iter} \) is somewhere in between these two extremes. To capture the overlap between \( T_{grad} \) and \( T_{sync} \), we model \( T_{iter} \) as
\begin{equation}
    T_{iter}(a, m, 0) = \left( T_{grad}(a, m)^\gamma + T_{sync}(a)^\gamma \right)^{1/\gamma},
    \label{eqn:overlap}
\end{equation}
where \( \gamma \ge 1 \) is a learnable parameter. Eqn.~\ref{eqn:overlap} has the property that \( T_{iter} = T_{grad} + T_{sync} \) when \( \gamma = 1 \), and smoothly transitions towards \( T_{iter} = \max(T_{grad}, T_{sync}) \) as \( \gamma \to \infty \).


\noindent\textbf{Gradient Accumulation.} In data-parallelism, GPU memory limits the per-GPU batch size, and many DL models hit this limit before the batch size is large enough for \( T_{grad} \) to overcome \( T_{sync} \) (or experience diminishing statistical efficiency), resulting in suboptimal scalability. Several techniques exist for overcoming the GPU memory limit~\cite{DBLP:journals/corr/ChenXZG16, 10.1145/2901318.2901323, 10.1145/3373376.3378530, MLSYS2020_084b6fbb}; we focus on gradient accumulation, which is easily implemented using popular DL frameworks. Per-GPU gradients are aggregated locally over \( s \) forward-backward passes before being synchronized across all GPUs during the \( (s+1)^{\text{th}} \) pass, achieving a larger total batch size. Thus, one iteration of SGD spans \( s \) accumulation steps followed by one synchronization step, modeled as
\begin{equation}
    \small
    T_{iter}(a, m, s) = s \times T_{grad}\left(a, m\right)
    + \left( T_{grad}\left(a, m\right)^\gamma + T_{sync}(a)^\gamma \right)^{1/\gamma}.
    \label{eqn:accumulation}
\end{equation}

\noindent\textbf{Throughput model validation.} 
Fig.~\ref{fig:throughput-model} shows an example of our \( \mathtt{THROUGHPUT} \) function fit to measured throughput values for a range of resource allocations and batch sizes. Each DL task was implemented using PyTorch~\cite{NIPS2019_9015}, which overlaps the backward pass' computation and communication. Gradients are synchronized with NCCL 2.7.8, which uses either ring all-reduce or tree all-reduce depending on the detected GPUs and their placements and its own internal performance estimates.
Overall, we find that our model can represent the observed data closely, while varying both the amount of resources as well as the batch size. In particular, all models we measured except ImageNet exhibited high sensitivity to inter-node synchronization, indicating that they benefit from co-location of GPUs. Furthermore, YOLOv3 and BERT benefit from using gradient accumulation to increase their total batch sizes. These detailed characteristics are well-represented by our \( \mathtt{THROUGHPUT} \) function, and can be optimized for by \oursystem{}.

In addition to the configurations in Fig.~\ref{fig:throughput-model}, we fitted the \( \mathtt{THROUGHPUT} \) function on a diverse set of GPU placements and batch sizes in a 64-GPU cluster. Across all DL tasks, the average error of the fitted model was at most 10\%, indicating that it represents the observed throughput measurements well.

\noindent\textbf{Limits of the throughput model.} \oursystem{} models data-parallel training throughput only in the dimensions it cares about, i.e. number and co-locality of GPUs, batch size, and gradient accumulation steps. The simple linear assumptions made in Eqn.~\ref{eqn:accumulation}, although sufficiently accurate for the settings we tested, may diverge from reality for specialized hardware~\cite{Jouppi2017IndatacenterPA}, sophisticated synchronization algorithms~\cite{Canny2013ButterflyMA,Zhao2014KylixAS,wei2015managed}, different parallelization strategies~\cite{NEURIPS2019_093f65e0,10.1145/3341301.3359646,Shoeybi2019MegatronLMTM,NEURIPS2018_3a37abde}, at larger scales~\cite{NEURIPS2020_1457c0d6,yamazaki2019accelerated}, or hidden resource contention not related to network used for gradient synchronization. Rather than attempting to cover all scenarios with a single throughput model, we designed \( \mathtt{GOODPUT}_t \) (Eqn.~\ref{eqn:goodput}) to be modular so that different equations for \( \mathtt{THROUGHPUT} \) may be easily plugged in without interfering with the core functionalities provided by \oursystem{}.


\section{\oursystem{} Design and Architecture}

%

\begin{figure}
\centering
\begin{subfigure}{.48\textwidth}
  \centering
  \includegraphics[page=1, trim=130 95 170 80, clip, width=\textwidth]{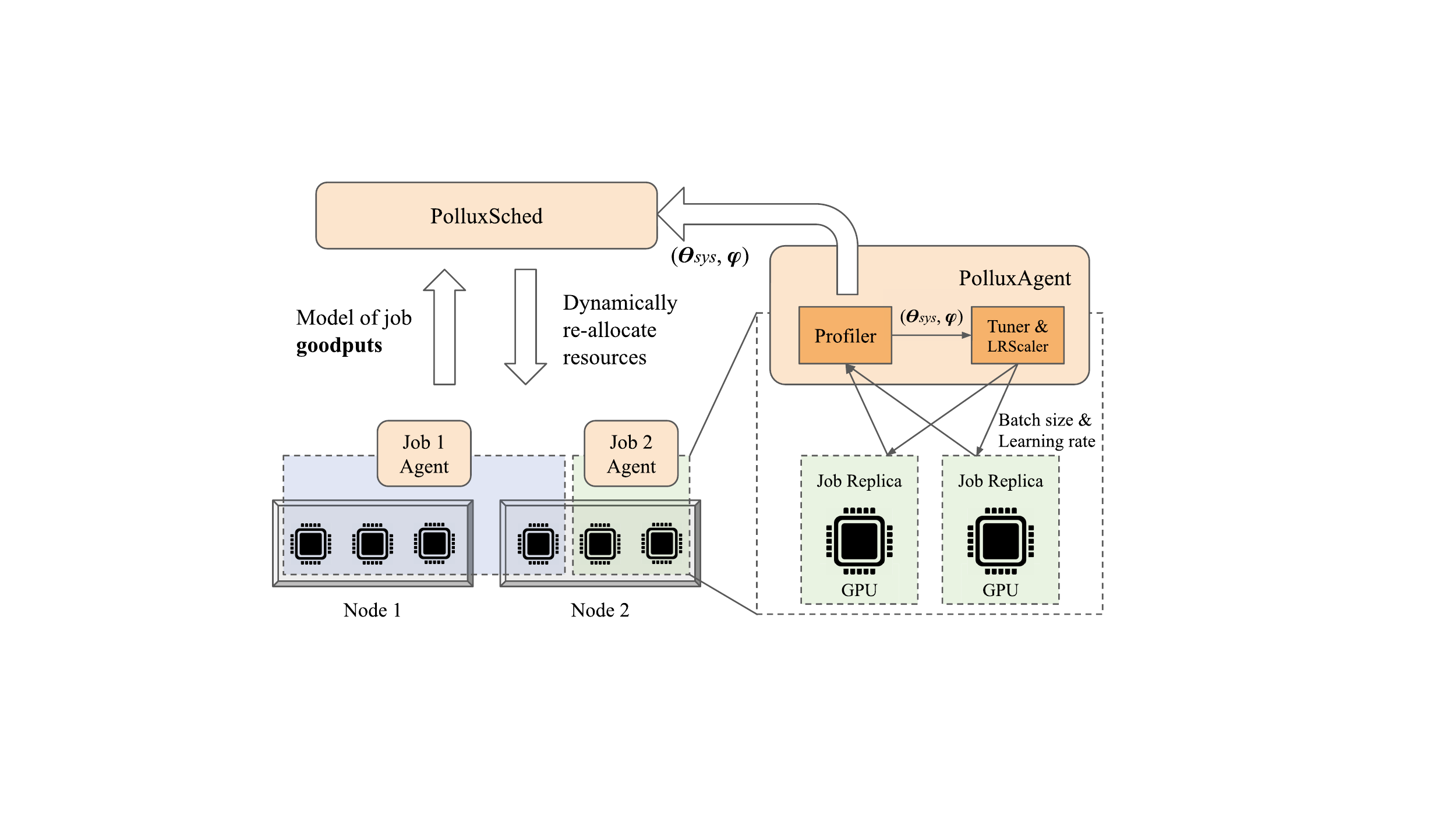}
\end{subfigure}
\caption{Co-adaptive scheduling architecture of \oursystem{}.}
\label{fig:scheduler-architecture}
\end{figure}

\oursystem{} adapts DL job execution at two distinct granularities. 
First, at a job-level granularity, \oursystem{} dynamically tunes the batch size and learning rate for best utilization of the allocated resources. 
Second, at the cluster-wide granularity, \oursystem{} dynamically (re-)allocates resources, driven by the goodput of all jobs sharing the cluster combined with cluster-level goals including fairness and job-completion time. 
To achieve this co-adaptivity in a scalable way, \oursystem{}'s design consists of two primary components, as illustrated in Fig.~\ref{fig:scheduler-architecture}.

First, a \emph{\oursystem{}Agent} runs together with each job. It fits the \( \mathtt{EFFICIENCY}_t \) and \( \mathtt{THROUGHPUT} \) functions for that job, and tunes its batch size and learning rate for efficient utilization of its current allocated resources. \oursystem{}Agent periodically reports the goodput function of its job to the \oursystem{}Sched.

Second, the \emph{\oursystem{}Sched} periodically optimizes the resource allocations for all jobs in the cluster, taking into account the current goodput function for each job and cluster-wide resource contention. Scheduling decisions made by \oursystem{}Sched also account for the overhead associated with resource re-allocations, slowdowns due to network interference between multiple jobs, and resource fairness.

\oursystem{}Agent and \oursystem{}Sched \emph{co-adapt} to each other. While \oursystem{}Agent adapts each training job to make efficient use of its allocated resources, \oursystem{}Sched dynamically re-allocates each job's resources, taking into account the \oursystem{}Agent's ability to tune its job.

\subsection{\oursystem{}Agent: Job-level Optimization}
\label{sec:agent}

An instance of \oursystem{}Agent is started with each training job. During training, it continually measures the job's gradient noise scale and system throughput, and it reports them to \oursystem{}Sched at a fixed interval. It also uses this information to determine the most efficient batch size for its job given its current resource allocations, and adapts its job's learning rate to this batch size using the appropriate plug-in LR scaling rule (e.g. AdaScale for SGD or square-root scaling for Adam).

\noindent\textbf{Online model fitting.}
In \S\ref{sec:modeling-system-throughput}, we defined the system throughput parameters of a training job as the 7-tuple
\begin{equation}
    \theta_{sys} = \left(\alpha_{grad}, \beta_{grad}, \alpha_{sync}^{local}, \beta_{sync}^{local}, \alpha_{sync}^{node}, \beta_{sync}^{node}, \gamma\right),
\end{equation}
which are required to construct the \( \mathtt{THROUGHPUT} \) function. Together with the PGNS \( \varphi_t \) (for predicting $\mathtt{EFFICIENCY}_t$) and initial batch size \( M_0 \), the triple \( ( \theta_{sys}, \varphi_t, M_0 ) \) specifies the \( \mathtt{GOODPUT} \) function. While \( M_0 \) is a constant configuration provided by the user, and \( \varphi_t \) can be computed according to \S\ref{sec:modeling-statistical-efficiency}, \( \theta_{sys} \) is estimated by fitting the \( \mathtt{THROUGHPUT} \) function to observed throughput values collected about the job during training.

\oursystem{}Agent measures the time taken per iteration, \( T_{iter} \), and records the tuple \( (a, m, s, T_{iter}) \) for all combinations of resource allocations \( a \), per-GPU batch size \( m \), and gradient accumulation steps \( s \) encountered during its lifetime.
Periodically, \oursystem{}Agent fits the parameters \( \theta_{sys} \) to all of the throughput data collected so far. Specifically, we minimize the root mean squared logarithmic error (RMSLE) between Eqn.~\ref{eqn:accumulation} and the collected data triples, using L-BFGS-B \cite{10.1145/279232.279236}. We set constraints for each \( \alpha \) and \( \beta \) parameter to be non-negative, and \( \gamma \) to be in the range \( [1, 10] \). \oursystem{}Agent then reports the updated values of \( \theta_{sys} \) and  \( \varphi_t \) to \oursystem{}Sched.

\noindent\textbf{Prior-driven exploration.}
At the beginning of each job, throughput values have not yet been collected.
To ensure that \oursystem{} finds efficient resource allocations through systematic exploration,
we impose several priors which bias \( \theta_{sys} \) towards
the belief that throughput scales perfectly with more resources, until such resource configurations are explored.

In particular, we set \( \alpha_{sync}^{local} = 0 \) while the job had not used more than one GPU, \( \alpha_{sync}^{local} = \beta_{sync}^{local} = 0 \) while the job had not used more than one node, and \( \beta_{sync}^{local} = \beta_{sync}^{node} = 0 \) while the job had not used more than two GPUs.
This creates the following behavior: each job starts with a single GPU and is initially assumed to scale perfectly to more GPUs. \oursystem{}Sched is then encouraged to allocate more GPUs and/or nodes to the job, naturally as part of its resource optimization (\S\ref{sec:sched}), until the \oursystem{}Agent can estimate \( \theta_{sys} \) more accurately. Finally, to prevent a job from being immediately scaled out to arbitrarily many GPUs, we restrict the maximum number of GPUs that can be allocated to at most twice the maximum number of GPUs the job has been allocated in its lifetime.

Although other principled approaches to exploration can be applied (e.g., Bayesian optimization), we find that this simple prior-driven strategy is sufficient in our experiments. Sec.~\ref{sec:scheduling-effects} shows that prior-driven exploration performs close (within 2-5\%) to an idealized scenario in which the model is fitted offline for each job before being submitted to the cluster.



\noindent\textbf{Training job tuning.}
With \( \theta_{sys} \), \( \varphi_t \), and \( M_0 \), which fully specify the DL job's \( \mathtt{GOODPUT} \) function at its current training progress, \oursystem{}Agent determines the most efficient per-GPU batch size and gradient accumulation steps,
\begin{equation}
    (m^*, \; s^*) = \arg\max_{m,s} \;\mathtt{GOODPUT}(a, m, s), \label{eqn:goodput_argmax}
\end{equation}
where \( a \) is the job's current resource allocation.

Once a new configuration is found, the job will use it for its subsequent training iterations, using the plug-in LR scaling rule to adapt its learning rate appropriately. As the job's \( \mathtt{EFFICIENCY}_t \) function changes over time, \oursystem{}Agent will periodically re-evaluate the most efficient configuration.

\subsection{\oursystem{}Sched: Cluster-wide Optimization}
\label{sec:sched}

The \oursystem{}Sched periodically allocates (and re-allocates) resources for every job in the cluster. To determine a set of efficient cluster-wide resource allocations, it maximizes a \emph{fitness function} that is defined as a generalized (power) mean across speedups for each job:
\begin{equation}
    \label{eqn:objective}
    \mathtt{FITNESS}_p(A) = \left(\frac{1}{J}\sum_{j=1}^J\mathtt{SPEEDUP}_{j}(A_j)^p\right)^{1/p}.
\end{equation}
\( A \) is an \textit{allocation matrix} with each row \( A_j \) being the allocation vector for a job \( j \), thus \( A_{jn} \) is the number of GPUs on node \( n \) allocated to job \( j \), and \( J \) is the total number of running and pending jobs sharing the cluster.
We define the speedup of each job as the factor of goodput improvement using a given resource allocation over using a fair-resource allocation, ie.
\begin{equation}
    \mathtt{SPEEDUP}_j(A_j) = \frac{\max_{m,s}\mathtt{GOODPUT}_j(A_j, m, s)}{\max_{m,s}\mathtt{GOODPUT}_j(a_f, m, s)},
\end{equation}
where \( \mathtt{GOODPUT}_j \) is the goodput of job \( j \) at its current training iteration, and \( a_f \) is a fair resource allocation for the job, defined to be an exclusive \( 1/J \) share of the cluster.\footnote{We note that \( \mathtt{SPEEDUP} \) has similarities with \textit{finish-time fairness}~\cite{themis2020}. But, \( \mathtt{SPEEDUP} \) is related to training performance at a moment in time, whereas finish-time fairness is related to end-to-end job completion time.}

In \S\ref{sec:goodput}, we described how the \( \mathtt{GOODPUT} \) function can be fitted to observed metrics during training and then be evaluated as a predictive model. \oursystem{}Sched leverages this ability to predict \( \mathtt{GOODPUT} \) to maximize \( \mathtt{FITNESS} \) via a search procedure, and then it applies the outputted allocations to the cluster.

\noindent\textbf{Fairness and the effect of \( p \).}
When \( p=1 \), \( \mathtt{FITNESS}_p \) is the average of \( \mathtt{SPEEDUP} \) values across all jobs.
This causes \oursystem{}Sched to allocate more GPUs to jobs that achieve a high $\mathtt{SPEEDUP}$ when provided with many GPUs (i.e., jobs that scale well). However, as \( p \to -\infty \), \( \mathtt{FITNESS}_p \) smoothly approaches the minimum of \( \mathtt{SPEEDUP} \) values, in which case maximizing \( \mathtt{FITNESS}_p \) promotes equal \( \mathtt{SPEEDUP} \) between training jobs, but ignores the overall cluster goodput and resource efficiency.


Thus, \( p \) can be considered a ``fairness knob'', with larger negative values being more fair. 
A cluster operator may select a suitable value, 
based on organizational priorities.
In our experience and results in \S\ref{sec:evaluation}, we find that \( p=-1 \) achieves 
most goodput improvements and reasonable fairness.

\noindent\textbf{Re-allocation penalty.} Each time a job is re-allocated to a different set of GPUs, it incurs some delay to re-configure the training process. Using the the popular checkpoint-restart method, we measured between 15 and 120 seconds of delay depending on the size of the model being trained and other initialization tasks in the training code. To prevent an excessive number of re-allocations, when \oursystem{}Sched evaluates the fitness function for a given allocation matrix, it applies a penalty for every job that needs to be re-allocated,
\begin{equation*}
    \mathtt{SPEEDUP}_j(A_j) \longleftarrow \mathtt{SPEEDUP}_j(A_j) \times \mathtt{REALLOC\_FACTOR}_j(\delta).
\end{equation*}

We define \( \mathtt{REALLOC\_FACTOR}_j(\delta) = (T_j - R_j \delta) / (T_j + \delta) \), where \( T_j \) is the age of the training job, \( R_j \) is the number of re-allocations incurred by the job so far, and \( \delta \) is an estimate of the re-allocation delay. Intuitively, \( \mathtt{REALLOC\_FACTOR}_j(\delta) \) scales \( \mathtt{SPEEDUP}_j(A_j) \) according to the assumption that the historical average rate of re-allocations for job \( j \) will continue indefinitely into the future. Thus, a job that has historically experienced a higher rate of re-allocations will be penalized more for future re-allocations.

\noindent\textbf{Interference avoidance.}
When multiple distributed DL jobs share a single node, their network usage while synchronizing gradients and model parameters may interfere with each other, causing both jobs to slow down~\cite{234916}; Xiao et al.~\cite{222611} report up to 50\% slowdown for DL jobs which compete with each other for network resources. \oursystem{}Sched mitigates this issue by disallowing different distributed jobs (each using GPUs across multiple nodes) from sharing the same node.

Interference avoidance is implemented as a constraint in \oursystem{}'s
search algorithm, by ensuring at most one distributed job is allocated to each node. We study the effects of interference avoidance in \S\ref{sec:scheduling-effects}.

\noindent\textbf{Supporting non-adaptive jobs.} In certain cases, a user may want to run a job with a fixed batch size, i.e. \( M = M_0 \). These jobs are well-supported by \oursystem{}Sched, which simply fixes \( \mathtt{EFFICIENCY}_t \) for that job to~1 and can continue to adapt its resource allocations based solely on its system throughput.

\subsection{Implementation}

\oursystem{}Agent is implemented as a Python library that is imported into DL training code.
We integrated \oursystem{}Agent with PyTorch~\cite{NIPS2019_9015}, which uses all-reduce as its gradient synchronization algorithm.
\oursystem{}Agent inserts performance profiling code that measures the time taken for each iteration of training, as well as calculating the gradient noise scale.
At a fixed time interval, \oursystem{}Agent fits the system throughput model (Eqn.~\ref{eqn:overlap}) to the profiled metrics collected so far, and reports the fitted system throughput parameters, along with the latest gradient statistics, to \oursystem{}Sched.
After reporting to \oursystem{}Sched, \oursystem{}Agent updates the job's per-GPU batch size and gradient accumulation steps, by optimizing its now up-to-date goodput function (Eqn.~\ref{eqn:goodput}) with its currently allocated resources.



\oursystem{}Sched is implemented as a service in Kubernetes~\cite{Kubernetes}. At a fixed time interval, \oursystem{}Sched runs
its search algorithm, and then applies the resultant allocation matrix by creating and terminating Kubernetes Pods that run the job workers.
To find a good allocation matrix, \oursystem{}Sched uses a population-based search algorithm that perturbs and combines candidate allocation matrices to produce higher-value allocation matrices, and finally modifies them to satisfy node resource constraints and interference avoidance. The allocation matrix with the highest fitness score is applied to the jobs running in the cluster.

Both \oursystem{}Agent and \oursystem{Sched} require a sub-procedure that optimizes \( \mathtt{GOODPUT}_t(a, m, s) \) given a fixed \( a \) (Eqn.~\ref{eqn:goodput_argmax}). We implemented this procedure by first sampling a range of candidate values for the total batch size \( M \), then finding the smallest \( s \) such that \( m = \left\lceil M / s \right\rceil \) fits into GPU memory according to a user-defined upper-bound, and finally taking the configuration which results in the highest \( \mathtt{GOODPUT} \) value.

\section{Evaluation}
\label{sec:evaluation}

We compare \oursystem{} with two state-of-the-art DL schedulers using a  testbed cluster with 64 GPUs. Although one primary advantage of \oursystem{} is automatically selecting the configurations for each job, we find that \oursystem{} still reduces average job completion times by \( 37 \)--\( 50 \)\% even when the baseline schedulers are supplied with well-tuned job configurations (a scenario that strongly favors the baseline schedulers). \oursystem{} is able to dynamically adapt each job by trading-off between high-throughput/low-efficiency and low-throughput/high-efficiency modes of training, depending on the current cluster state and training progress.

Using a cluster simulator, we evaluate the impact of specific settings on \oursystem{}, including the total workload intensity, prior-driven exploration, scheduling interval, and interference avoidance. With its fairness knob, \oursystem{} can improve finish-time fairness~\cite{themis2020} by \(1.5\)--\(5.4\times\) compared to baseline DL schedulers. We also reveal a new opportunity for auto-scaling in the cloud by showing that a \oursystem{}-based auto-scaler can potentially reduce the cost of training large models (e.g. ImageNet) by 25\%.

\subsection{Experimental Setup}
\label{sec:experimental-setup}

\noindent\textbf{Testbed.} We conduct experiments using a cluster consisting of 16 nodes and 64 GPUs. Each node is an AWS EC2 \texttt{g4dn.12xlarge} instance with 4 NVIDIA T4 GPUs, 48 vCPUs, 192GB memory, and a 900GB SSD. All instances are launched within the same placement group. We deployed Kubernetes 1.18.2 on this cluster, along with CephFS 14.2.8 to store checkpoints for checkpoint-restart elasticity.

\noindent\textbf{Synthetic Workload Construction.} 
We randomly sampled 160 jobs from the busiest 8-hour range (hours 3--10) in the deep learning cluster traces published by Microsoft~\cite{234916}.
Each job in the orginal trace has information on its submission time, number of GPUs, and duration. However, no information is provided on the model architectures being trained or dataset characteristics. Instead, our synthetic workload consists of the models and datasets described in Table~\ref{tab:models}.


\begin{table*}
\centering
\footnotesize
\begin{tabular}{|c|c|c|c|c|c|c|c|c|} 
  \hline
  \textbf{Task} & \textbf{Dataset} & \textbf{Model} & \textbf{Optimizer} & \textbf{LR Scaler} & \( \mathbf{M_0} \) & \textbf{Validation} & \textbf{Size} & \textbf{Frac. Jobs} \\
  \hline\hline
  Image Classification & ImageNet~\cite{deng2009imagenet} & ResNet-50~\cite{7780459} & SGD & AdaScale & 200 imgs & 75\% top1 acc. & XL & 2\% \\
  \hline
  Object Detection & PASCAL-VOC~\cite{Everingham10} & YOLOv3~\cite{DBLP:journals/corr/abs-1804-02767} & SGD & AdaScale & 8 imgs & 84\% mAP & L & 6\% \\
  \hline
  Speech Recognition & CMU-ARCTIC~\cite{cmuarctic} & DeepSpeech2~\cite{10.5555/3045390.3045410} & SGD & AdaScale & 20 seqs & 25\% word err. & M & 10\% \\
  \hline
  Question Answering & SQuAD~\cite{rajpurkar2016squad} & BERT (finetune)~\cite{devlin2019bert} & AdamW & Square-Root &  12 seqs &  88\% F1 score & M & 10\% \\
  \hline
  Image Classification & Cifar10~\cite{cifar10} & ResNet18~\cite{7780459} & SGD & AdaScale & 128 imgs & 94\% top1 acc. & S & 36\% \\
  \hline
  Recommendation & MovieLens~\cite{10.1145/2827872} & NeuMF~\cite{10.1145/3038912.3052569} & Adam & Square-Root & 256 pairs & 69\% hit rate & S & 36\% \\
  \hline
\end{tabular}
\caption{Models and datasets used in our evaluation workload. Each training task achieves the provided validation metrics. The fraction of jobs from each category are chosen according to the public Microsoft cluster traces.}
\label{tab:models}
\end{table*}

We categorized each job in the trace and in Table~\ref{tab:models} based on their total GPU-time: Small (0--1 GPU-hours), Medium (1--10 GPU-hours), Large (10--100 GPU-hours), and XLarge (100--1000 GPU-hours). 
For each job in the trace, we picked a training job from Table~\ref{tab:models} that is in the same category.

\noindent\textbf{Manually-tuned jobs for baseline DL schedulers.} 
We manually tuned the number of GPUs and batch sizes for each job in our synthetic workload, as follows. We measured the time per training iteration for each model in Table~\ref{tab:models} using a range of GPU allocations and batch sizes, and fully trained each model using a range of different batch sizes (see \S\ref{sec:simulator-experiments} for details). We considered a number of GPUs \emph{valid} if using the optimal batch size for that number of GPUs achieves 50\% -- 80\% of the ideal (i.e., perfectly linear) scalability versus using the optimal batch size on a single GPU. For each job submitted from our synthetic workload, we selected its number of GPUs and batch size randomly from its set of valid configurations.

Our job configurations assume that the users are highly rational and knowledgeable about the scalability of the models they are training. Less than \( 50\% \) of the ideal scalability would lead to under-utilization of resources, and more than \( 80\% \) of the ideal scalability means the job can still utilize more GPUs efficiently. We emphasize that this assumption of uniformly sophisticated users is unrealistically biased in favor of the baseline schedulers and only serves for comparing \oursystem{} with the ideal performance of baseline systems.


\noindent\textbf{Comparison of DL schedulers.} We compare \oursystem{} to two recent deep learning schedulers, Tiresias~\cite{227623} and Optimus~\cite{10.1145/3190508.3190517}, as described in \S\ref{sec:existing-schedulers}.
Whereas \oursystem{} dynamically co-adapts the number of GPUs and batch sizes of DL training jobs, Optimus only adapts the number of GPUs, and Tiresias adapts neither. To establish a fair baseline for comparison, for all three schedulers, we scale the learning rate using AdaScale for SGD, and the square-root scaling rule for Adam and AdamW.

\textit{\oursystem{}.} We configured \oursystem{}Sched to use a 60s scheduling interval, and compute \( \mathtt{REALLOC\_FACTOR(\delta)} \) using \( \delta = 30\text{s} \). \oursystem{}Agent reports its most up-to-date system throughput parameters and gradient statistics every 30s.
Unless otherwise specified, the default fairness knob value of \( p = -1 \) is used.

\textit{Tiresias.} We configured Tiresias as described in the testbed experiments of Gu et al.~\cite{227623}, with two priority queues and the \texttt{PromoteKnob} disabled. We manually tuned the queue threshold to perform well for our synthetic workload. Whenever possible, we placed jobs onto as few different nodes as possible to promote worker locality.

\textit{Optimus+Oracle.} Optimus leverages a throughput prediction model that is specific to jobs using the parameter server architecture. To account for differences due to the performance model, our implementation of Optimus uses our own throughput model as described in \S\ref{sec:modeling-system-throughput}. Furthermore, Optimus predicts the number of training iterations until convergence by fitting a simple function to the model's convergence curve. Since this method does not work consistently for all models in our synthetic workload, we run each job ahead of time and provide Optimus with the exact number of iterations until completion. We call this version of Optimus \textit{Optimus+Oracle}.

For each job, Tiresias uses the number of GPUs and batch size specified in our synthetic workload. Optimus+Oracle uses the batch size specified, but determines the number of GPUs dynamically. Each job uses gradient accumulation if they are allocated too few GPUs to support the specified batch size.


\subsection{Testbed Macrobenchmark Experiments}
\label{sec:testbed-experiments}

Table~\ref{tab:testbed-results} summarizes the results of our testbed experiments for seven configurations: \oursystem{} compared with, first, baseline schedulers using well-tuned job configurations; second, baseline schedulers using more realistic job configurations; third, \oursystem{} using two alternate values for its fairness knob.

\begin{table}
\centering
\small
\begin{tabular}{|c|c|c|c|} 
  \hline
  \multirow{2}{*}{\textbf{Policy}} & \multicolumn{2}{|c|}{\footnotesize{\textbf{Job Completion Time}}} & \multirow{2}{*}{\textbf{Makespan}} \\
  \cline{2-3}
  & \footnotesize{\textbf{Average}} & \footnotesize{\textbf{99\%tile}} & \\
  \hline\hline
  \oursystem{} (\( p=-1 \)) & \textbf{0.76h} & 11h & \textbf{16h} \\
  \hline
  \footnotesize{Optimus+Oracle+TunedJobs} & 1.5h & 15h & 20h \\
  \hline
  Tiresias+TunedJobs & 1.2h & 15h & 24h \\
  \hline \hline
  Optimus+Oracle & 2.7h & 22h & 28h \\
  \hline
  Tiresias & 2.8h & 25h & 31h \\
  \hline\hline
  \oursystem{} (\( p=+1 \)) & 0.83h & \textbf{10h} & \textbf{16h} \\
  \hline
  \oursystem{} (\( p=-10 \)) & 0.84h & 12h & 18h \\
  \hline
\end{tabular}
\caption{Summary of testbed experiments.
}
\label{tab:testbed-results}
\end{table}

\noindent\textbf{Comparisons using well-tuned job configurations.} Even when Optimus+Oracle and Tiresias are given well-tuned job configurations as described in \S\ref{sec:experimental-setup}, they are still significantly behind \oursystem{}. In this setting, \oursystem{} (with \( p = -1 \)) achieved 50\% and 37\% shorter average JCT, 27\% and 27\% shorter tail (99th percentile) JCT, and 20\% and 33\% shorter makespan, in comparison to Optimus+Oracle+TunedJobs and Tiresias+TunedJobs, respectively. As we previously noted, this setting highly favors the baseline schedulers, essentially mimicking users who possess expert knowledge about system throughput, statistical efficiency, and how their values change with respect to resource allocations and batch sizes.

\begin{figure}
\centering
\begin{subfigure}{.48\textwidth}
  \includegraphics[width=\textwidth,trim=10 35 10 15,clip]{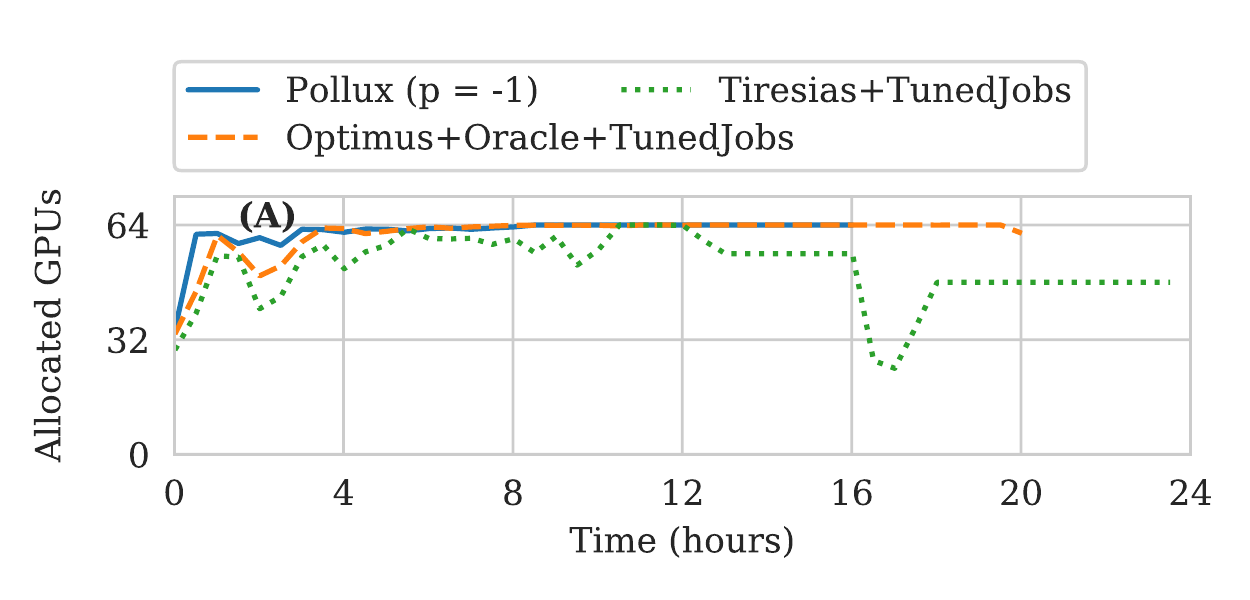}
\end{subfigure}
\begin{subfigure}{.48\textwidth}
  \includegraphics[width=\textwidth,trim=10 10 10 50,clip]{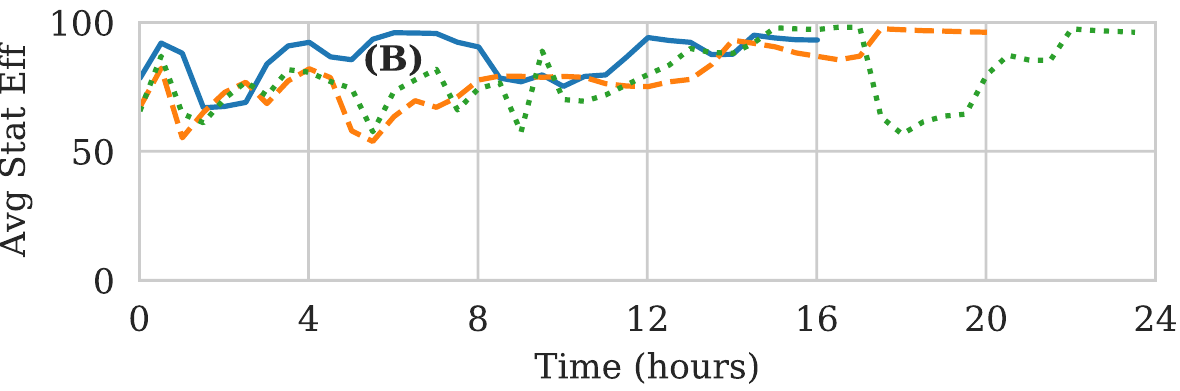}
\end{subfigure}
\caption{Comparison between \oursystem{} (\(p=-1\)), Optimus, and Tiresias while executing our synthetic workload (with tuned jobs). TOP: average cluster-wide allocated GPUs over time. BOTTOM: average cluster-wide statistical efficiency over time. Tiresias+TunedJobs dips between hours 16 and 20 due to a 24-GPU job blocking a 48-GPU job from running.}
\label{fig:physical-compare}
\end{figure}

One key source of improvement for \oursystem{} is its ability to trade-off between high-throughput/low-efficiency and low-throughput/high-efficiency modes during training.
Fig.~\ref{fig:physical-compare} shows the total number of allocated GPUs and average \( \mathtt{EFFICIENCY}_t \) during the execution of our synthetic workload.
During periods of low cluster contention, \oursystem{} can allocate more GPUs (indicated by \textbf{(A)}) and use larger batch sizes to boost training throughput, even at the cost of lower statistical efficiency, because doing so results in an overall higher goodput. On the other hand, during periods of high cluster contention, \oursystem{} may instead use smaller batch sizes to increase statistical efficiency (indicated by \textbf{(B)}).

\noindent\textbf{Comparisons using realistic job configurations.} Without assistance from a system like \oursystem{}, users are likely to try various numbers of GPUs and batch sizes, before finding a configuration that is efficient. Other users may not invest time into configuring their jobs well in the first place.

To set a more realistically configured baseline, we ran Optimus+Oracle and Tiresias on a version of our synthetic workload with the number of GPUs exactly as specified in the Microsoft cluster trace. The batch size was chosen to be the baseline batch size \( M_0 \) times the number of GPUs, which is how we expect most users to initially configure their distributed training jobs. We find that these jobs typically use fewer GPUs and smaller batch sizes than their well-configured counterparts.

Using this workload, we find that \oursystem{} has 72\% and 73\% shorter average JCT, 50\% and 56\% shorter tail JCT, and 43\% and 48\% shorter makespan, in comparison to Optimus+Oracle and Tiresias, respectively. Even though Optimus+Oracle can dynamically increase the GPU allocation of each job, it still only slightly outperforms Tiresias because it does not also increase the batch size to better utilize those additional GPUs.

\begin{figure}
\centering
\begin{subfigure}{.25\textwidth}
  \includegraphics[width=\textwidth,trim=10 35 0 10,clip]{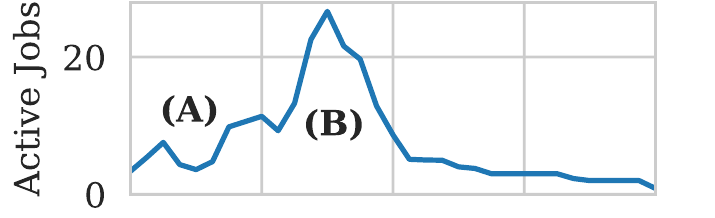}
\end{subfigure}%
\begin{subfigure}{.23\textwidth}
  \includegraphics[width=\textwidth,trim=24 36 0 10,clip]{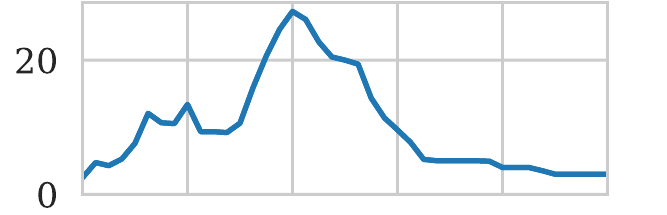}
\end{subfigure}
\begin{subfigure}{.25\textwidth}
  \includegraphics[width=\textwidth,trim=10 37 0 10,clip]{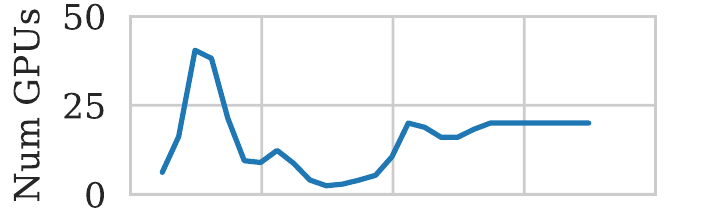}
\end{subfigure}%
\begin{subfigure}{.23\textwidth}
  \includegraphics[width=\textwidth,trim=24 37 0 10,clip]{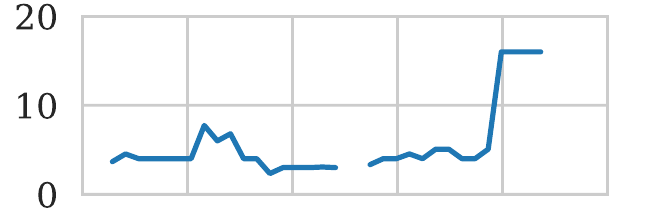}
\end{subfigure}
\begin{subfigure}{.25\textwidth}
  \includegraphics[width=\textwidth,trim=10 35 0 10,clip]{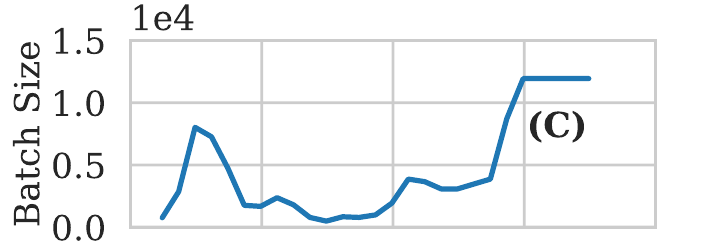}
\end{subfigure}%
\begin{subfigure}{.23\textwidth}
  \includegraphics[width=\textwidth,trim=24 37 0 10,clip]{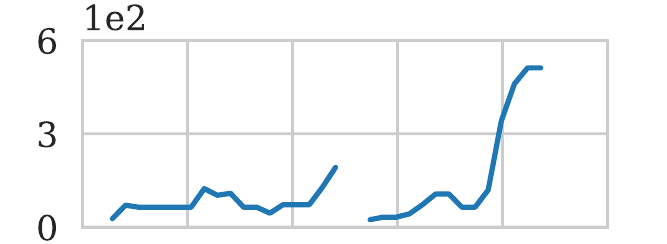}
\end{subfigure}
\begin{subfigure}{.25\textwidth}
  \includegraphics[width=\textwidth,trim=10 10 0 10,clip]{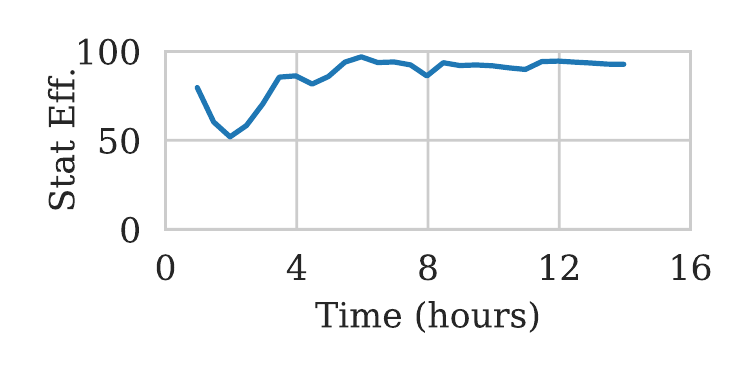}
\end{subfigure}%
\begin{subfigure}{.23\textwidth}
  \includegraphics[width=\textwidth,trim=24 10 0 10,clip]{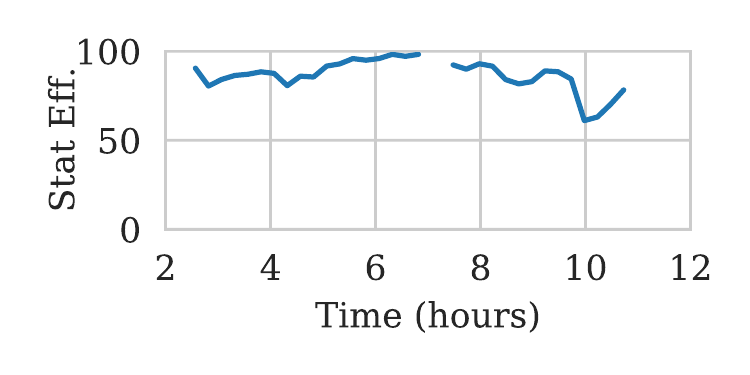}
\end{subfigure}
\caption{Co-adaptation over time of one ImageNet job (LEFT) and two YOLOv3 jobs (RIGHT) using \oursystem{} (\(p=-1\)). ROW 1: number of jobs actively sharing the cluster. ROW 2: number of GPUs allocated to the job. ROW 3: batch size (images) used. ROW 4: statistical efficiency (\%).}
\label{fig:physical-imagenet}
\end{figure}

\noindent\textbf{A closer look at co-adapted job configurations.} Fig.~\ref{fig:physical-imagenet} (LEFT) shows the configurations chosen by \oursystem{} for one ImageNet training job as the synthetic workload progresses. \textbf{(A)} during the initial period of low cluster contention, more GPUs are allocated to ImageNet, causing a larger batch size to be used and lowering statistical efficiency. \textbf{(B)} during the subsequent period of high cluster contention, fewer GPUs are allocated to ImageNet, causing a smaller batch size to be used and raising statistical efficiency. \textbf{(C)} when the cluster contention comes back down, ImageNet continues to be allocated more GPUs and uses a larger batch size. However, we note that the batch size per GPU is much higher than in the first low-contention period, since the job is now in its final, high-statistical-efficiency phase of training. We see similar trade-offs being made over time for two YOLOv3 jobs (RIGHT).

\noindent\textbf{Effect of the fairness knob.} We ran \oursystem{} using three values of the fairness knob, \( p = 1, -1, -10\). Compared with no fairness (\( p = 1 \)), introducing a moderate degree of fairness (\(p=-1\)) improved the average job completion time (JCT) but degraded the tail JCT. This is because\footnote{We note that \( p = -1 \) (harmonic mean over speedups) may be more suitable than \( p = 1 \) (arithmetic mean) when optimizing for the average JCT.}, in our synthetic workload, the tail JCT comprises of long but scalable jobs (i.e. ImageNet), which take a large number of GPUs away from other jobs in the absence of fairness (\(p=1\)). However, further increasing fairness (\(p = -10\)) degraded performance in average JCT, tail JCT, and makespan. In \S\ref{sec:eval-fairness}, we present a more detailed analysis of the impact of \( p \) on scheduling fairness.

\noindent\textbf{System overheads.} During each 60s scheduling interval, \oursystem{}Sched spent an average of 1~second on 1~vCPU computing the cluster allocations by optimizing the \( \mathtt{FITNESS}_p \) function. On average, each job was re-allocated resources once every 7 minutes, resulting in an average 8\% run-time overhead due to checkpoint-restarts. Each \oursystem{}Agent fits its throughput model parameters on its latest observed metrics every 30 seconds, taking an average of 0.2 seconds each time. Finding the optimal per-GPU batch size and gradient accumulation steps by optimizing \( \mathtt{GOODPUT}_t \) takes an average of 0.4 milliseconds.

\subsection{Simulator Experiments}
\label{sec:simulator-experiments}


We built a discrete-time cluster simulator in order to evaluate a broader set of workloads and settings.
Our simulator is constructed by measuring the performance and gradient statistics of each model in Table~\ref{tab:models}, under many different resource and batch size configurations, and re-playing them for each simulated job. This way, we are able to simulate both the system throughput and statistical efficiency of the jobs in our workload.

Unless stated otherwise, each experiment in this section is repeated on 8 different workload traces generated using the same duration, number of jobs, and job size distributions as in \S\ref{sec:testbed-experiments}, and we report the average results across all 8 traces.


\noindent\textbf{Simulator construction.} For each job in Table~\ref{tab:models}, we measured the time per training iteration for 146 different GPU allocations+placements in our testbed cluster of 16 nodes and 64 total GPUs. For each allocation, we measured a range of batch sizes up to the GPU memory limit. To simulate the throughput for a job, we queried a multi-dimensional linear interpolation on the configurations we measured. For each model, we also measured the (pre-conditioned) gradient noise scale during training using a range of batch sizes, and across every epoch. To simulate the statistical efficiency for a job using a certain batch size, we linearly interpolated its value of the PGNS between the two nearest batch sizes we measured.

\noindent\textbf{Simulator fidelity.} The data we collected about each job enables our simulator to reproduce several system effects, including the performance impact of different GPU placements. We also simulate the overhead of checkpoint-restarts by injecting a 30-second delay for each job that has its resources re-allocated. Unless stated otherwise, we do not simulate any network interference between different jobs. We study the effects of interference in more detail in \S\ref{sec:scheduling-effects}. 

Compared with our testbed experiments in \S\ref{sec:testbed-experiments}, we find that our simulator obtains similar factors of improvement, showing that \oursystem{} reduces the average JCT by 48\% and 32\% over Optimus+Oracle+TunedJobs and Tiresias+TunedJobs.

\subsubsection{Scheduling Fairness}
\label{sec:eval-fairness}

We evaluate the scheduling fairness of \oursystem{} using \emph{finish-time fairness}\cite{themis2020} (denoted by \( \rho \)), which is defined to be the ratio of a job's JCT running on shared resources to that of the job running in an isolated and equally-partitioned cluster. Under this metric, jobs with $\rho < 1$ have been treated better-than-fair by the cluster scheduler, while jobs with $\rho > 1$ have been treated worse-than-fair.

\begin{figure}
  \centering
  \includegraphics[width=0.4\textwidth]{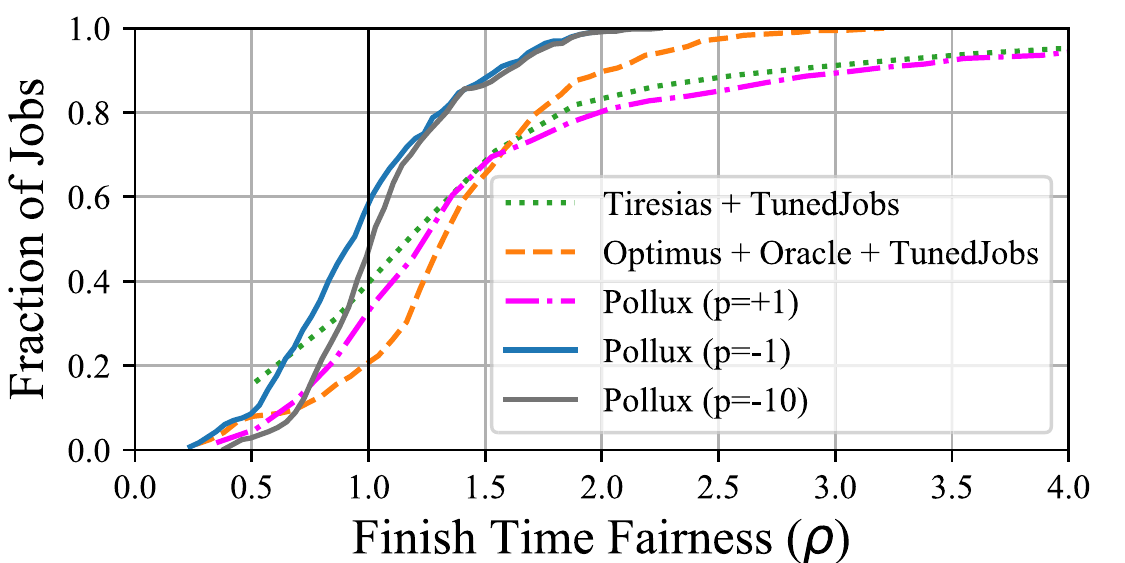}
  \caption{CDF of Finish Time Fairness ($\rho$).
  }
  \label{fig:finish-time-fairness}
\end{figure}

In Fig.~\ref{fig:finish-time-fairness}, we compare the finish-time fairness of \oursystem{} with Optimus+Oracle+TunedJobs and Tiresias+TunedJobs. \oursystem{} with \( p = 1 \) results in poor fairness, similar to Tiresias+TunedJobs, which is apparent as a long tail of jobs with \( \rho > 4 \). Optimus+Oracle+TunedJobs obtains better fairness due to its allocation algorithm which attempts to equalize the JCT improvement for each job. 
\oursystem{} with \( p = -1 \) provides the best fairness, 
with 99\% of jobs achieving \( \rho < 2 \), and does so while still providing significant performance increases (Table~\ref{tab:testbed-results}). 
For \( p = -10 \), we observe slightly worse fairness overall,
caused by
\oursystem{}Sched incurring a larger number of re-allocations due to ignoring the cost in favor of equalizing speedups at all times.

To provide context, we note that the curves for Tiresias and Optimus are consistent with those reported (for different workloads) by Mahajan et al.~\cite{themis2020}.
Although their Themis system is not available for direct comparison, the $\rho$~range for \oursystem{} with \( p = -1 \) is similar to the range reported for Themis.  
The max-$\rho$ improvements (1.5$\times$ and 5.4$\times$) over Tiresias and Optimus are also similar.



\subsubsection{Other Effects on Scheduling}
\label{sec:scheduling-effects}

\begin{figure}
\centering
\begin{subfigure}{.48\textwidth}
  \centering
  \includegraphics[width=\textwidth, trim=10 10 40 10, clip]{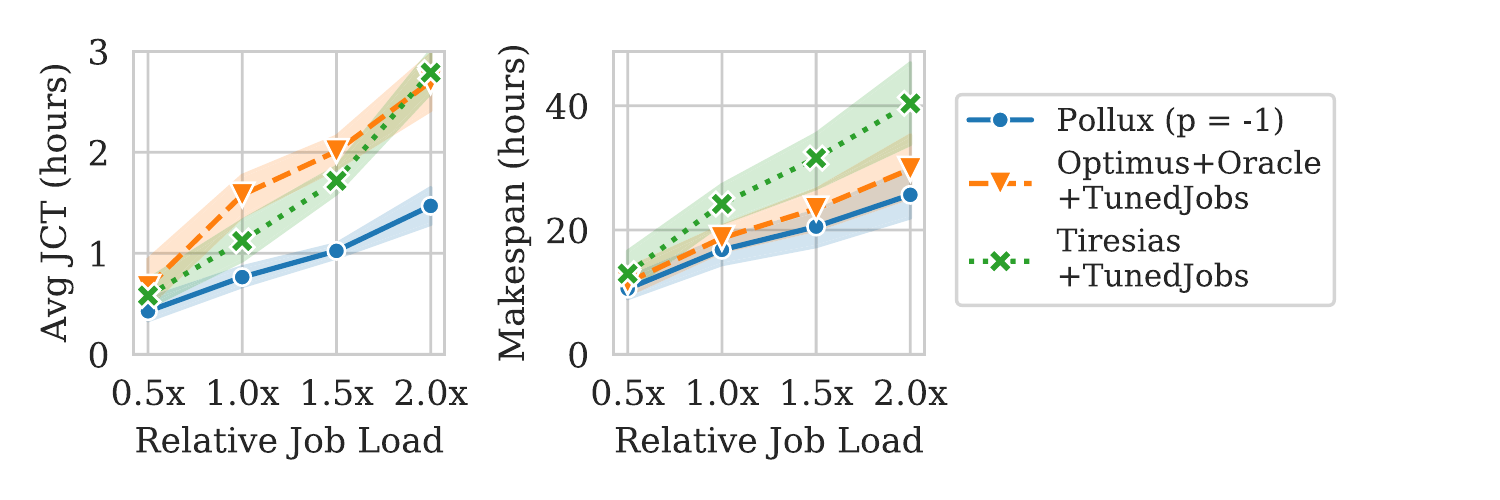}
  \caption{Varying the workload intensity.}
  \label{fig:workload-numjobs}
\end{subfigure}
\begin{subfigure}{.24\textwidth}
  \centering
  \includegraphics[width=\textwidth, trim=5 10 0 0, clip]{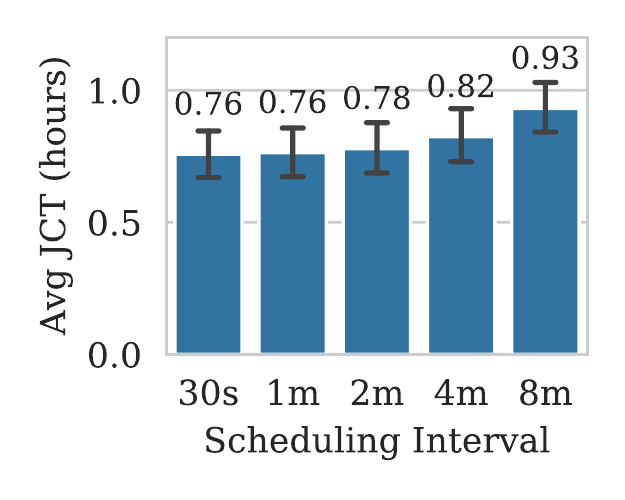}
  \caption{Varying scheduling interval.}
  \label{fig:scheduling-interval}
\end{subfigure}%
\begin{subfigure}{.25\textwidth}
  \centering
  \includegraphics[width=\textwidth,clip,trim=0 12 0 10]{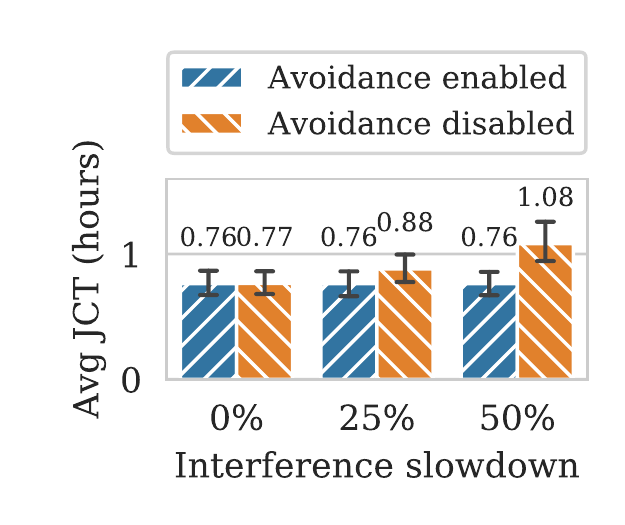}
  \caption{Varying job interference.}
  \label{fig:interference}
\end{subfigure}
\caption{Effects of various parameters on \oursystem{}, error bars and bands represent 95\% confidence intervals.}
\end{figure}

\noindent\textbf{Sensitivity to job load.} We compare the performance of \oursystem{}, Optimus+Oracle+TunedJobs, and Tiresias+TunedJobs for increasing workload intensity in terms of rate of job submissions. Fig.~\ref{fig:workload-numjobs} shows the results. As expected, all three scheduling policies suffer longer average JCT and makespan as the load is increased. Across all job loads, \oursystem{} maintains similar relative improvements over the baseline schedulers.

\noindent\textbf{Impact of prior-driven exploration.} 
\oursystem{} explores GPU allocations for each DL job from scratch during training (Sec.~\ref{sec:agent}). We evaluated the potential improvement from more efficient exploration by seeding each job's throughput models using historical data collected offline. We observed minor (2--5\%) reduction in JCT for short jobs like CIFAR10, but no significant change for longer running jobs, indicating low overhead from \oursystem{}'s prior-driven exploration.

\noindent\textbf{Impact of scheduling interval.} We ran \oursystem{} using a range of values for its scheduling interval, as shown in Fig.~\ref{fig:scheduling-interval}. We find that \oursystem{} performs similarly well in terms of average JCT for intervals up to 2 minutes, while longer intervals result in performance degradation. Since newly-submitted jobs can only start during the next scheduling interval, we would expect an increase in the average queuing time due to longer scheduling intervals. However, we find that queuing contributed to roughly half of the performance degradation observed, indicating that \oursystem{} still benefits from a relatively frequent adjustment of resource allocations.

\noindent\textbf{Impact of interference avoidance.} 
To evaluate the impact of \oursystem{}Sched's interference avoidance constraint, we artificially inject various degrees of slowdown for distributed jobs sharing the same node. Fig.~\ref{fig:interference} shows the results. With interference avoidance enabled, the average JCT is unaffected by even severe slowdowns, because network contention is completely mitigated. However, without interference avoidance, the average JCT is \( 1.4\times \) longer when the interference slowdown is \( 50\% \).
On the other hand, in the ideal scenario when there is zero slowdown due to interference, \oursystem{}Sched performs similarly whether or not interference avoidance is enabled. This indicates that \oursystem{}Sched is still able to find efficient cluster allocations while obeying the interference avoidance constraint.

\subsection{More Applications of \oursystem{}}

\subsubsection{Cloud Auto-scaling}
\label{sec:cloud-autoscaling}
%
%
In cloud environments, computing resources can be obtained and released as required, and users pay for the duration they hold onto those resources. 
Goodput-driven scheduling presents a unique opportunity: when a DL model's statistical efficiency increases during training, it may be more cost-effective to provision more cloud resources and use larger batch sizes during the later epochs of a large training job, rather than earlier on. We present some preliminary evidence using our cluster simulator, and note that a full design of an auto-scaling system based on goodput may be the subject of future work.

\noindent\textbf{Auto-scaling ImageNet training.}
We implemented a simple auto-scaling policy using \oursystem{}'s goodput function. During training, we scaled up the number of nodes whenever {\small\( \max_{m,s} \mathtt{GOODPUT}_t(a, m, s)/\mathtt{SUM}(a) > U\cdot \max_{m,s} \mathtt{GOODPUT}_t(1, m, s) \)}, i.e. the goodput exceeds some fraction \( U \) of the predicted ideal goodput assuming perfect scalability. We set \( U=2/3 \), and increased to a number of nodes such that the predicted goodput is approximately \( L=1/2 \) of the predicted ideal goodput.

\begin{figure}
\begin{subfigure}{.24\textwidth}
  \centering
  \includegraphics[width=\textwidth, trim=0 12 0 0, clip]{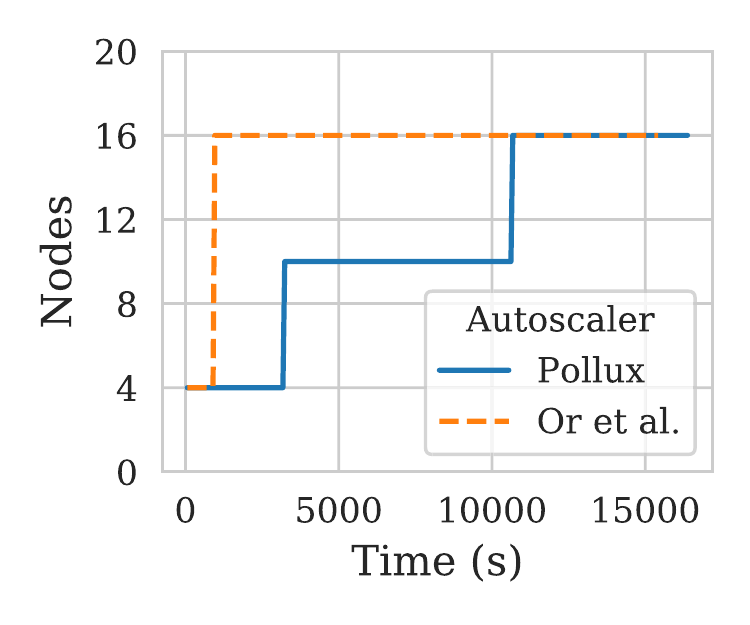}
  \caption{Number of nodes over time.}
  \label{fig:autoscale-imagenet-cluster}
\end{subfigure}\hspace*{\fill}%
\begin{subfigure}{.24\textwidth}
  \centering
  \includegraphics[width=\textwidth, trim=0 12 0 0, clip]{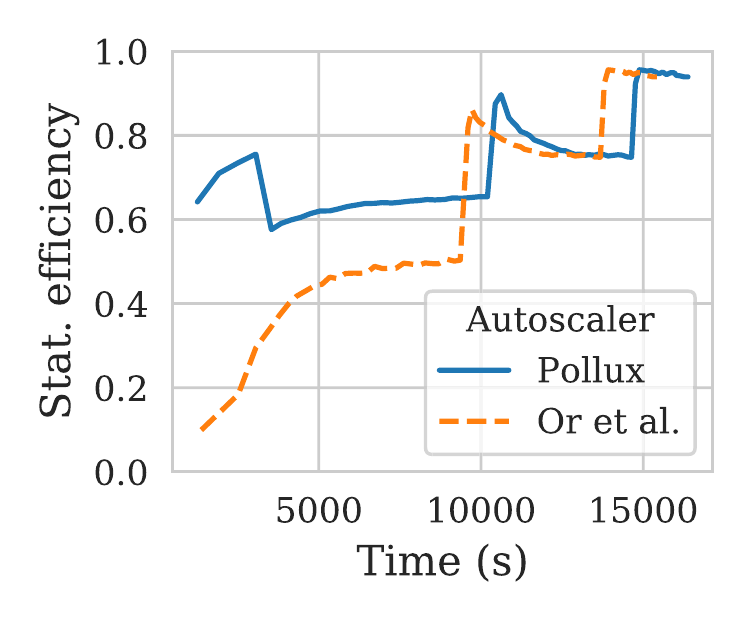}
  \caption{Statistical efficiency over time.}
  \label{fig:autoscale-imagenet-efficiency}
\end{subfigure}
\caption{Goodput-based auto-scaling (\oursystem{}) vs throughput-based auto-scaling (Or et al.) for ImageNet training.}
\label{fig:autoscale-imagenet}
\end{figure}

Fig.~\ref{fig:autoscale-imagenet} compares our \oursystem{}-based auto-scaler with the auto-scaler proposed by Or \emph{et al.}~\cite{mlsys2020_168}, which allows the batch size to be increased during training, but models job performance using the system throughput rather than the goodput. Since the system throughput does not change with training progress, throughput-based autoscaling (Or \emph{et al.}) quickly scales out to more nodes and a larger batch size (Fig.~\ref{fig:autoscale-imagenet-cluster}), which remains constant thereafter. On the other hand, \oursystem{} starts with a small number of nodes, and gradually increases the number of nodes as the effectiveness of larger batch sizes improves over time. Fig.~\ref{fig:autoscale-imagenet-efficiency} shows that \oursystem{} maintains a high statistical efficiency throughout training.
Overall, compared to Or \emph {et al.}'s throughput-based auto-scaling, \oursystem{} trains ImageNet with 25\% cheaper cost, with only a 6\% longer completion time.

\subsubsection{Hyper-parameter Optimization (HPO)}
%
Hyper-parameter optimization (HPO) 
is an
important DL workload.
In HPO, the user defines a \textit{search space} over relevant model hyper-parameters. 
A HPO algorithm (aka a trial scheduler) submits many training jobs (trials) to evaluate the effectiveness of particular hyper-parameters, in terms of objectives such as model accuracy or energy efficiency.

Different HPO algorithm types manage trials differently.
For example, Bayesian optimization algorithms~\cite{klein2017fast, snoek2012practical} may submit a few training jobs at a time, and determine future trials based on the fully-trained results of previous trials. 
Bandit-based algorithms~\cite{li2017hyperband} may launch a large number of trials at once and early-stop ones that appear unpromising.

A full evaluation on how \oursystem{} affects different HPO algorithm types is
future work.
Table~\ref{tab:hpo} shows results from 
%
tuning
a ResNet18 model trained on the CIFAR10 dataset, using a popular Bayesian optimization-based HPO algorithm known as the Tree-structured Parzen Estimator (TPE)~\cite{bergstra2011algorithms}.
The search space covers the learning rate and annealing, momentum, weight decay, and network width hyper-parameters. We configured TPE so that 4 trials run concurrently with each other, and 100 trials are run in total. The testbed consists of two NVIDIA DGX A100 nodes, each with 8 A100 GPUs. The baseline scheduler assigns a static allocation of 4 GPUs (all on the same node) to each trial and uses a fixed per-GPU batch size for every trial.
As expected, similar accuracy values are achieved, but 
\oursystem{} completes HPO 30\% faster due to adaptive (re-)allocation of resources as trials progress and adaptive batch sizes.


\begin{table}
\centering
\small
\begin{tabular}{|c|c|c|c|}
\hline
\textbf{Policy} & \textbf{Accuracy (Top 5 trials)} & \textbf{Avg JCT} & \textbf{Makespan} \\
\hline\hline
Pollux & \( 95.4 \pm 0.2 \) & 25min & 10h \\
\hline
Baseline & \( 95.5 \pm 0.3 \) & 34min & 14h \\
\hline
\end{tabular}
\caption{Summary of HPO experiments.}
\label{tab:hpo}
\end{table}


\subsection{Artifact}

We provide an artifact containing the full implementation of \oursystem{}, benchmark model implementations (Table~\ref{tab:models}), testbed experiment scripts (Sec.~\ref{sec:testbed-experiments}), cluster simulator implementation and results (Sec.~\ref{sec:simulator-experiments}), available at \url{https://github.com/petuum/adaptdl/tree/osdi21-artifact}.
The raw testbed experiment (Sec.~\ref{sec:testbed-experiments}) logs and analysis scripts are provided at \url{https://github.com/petuum/pollux-results}.

\section{Additional Related Work}
\label{sec:related-work}

Prior DL schedulers are discussed in \S\ref{sec:existing-schedulers}.

\noindent\textbf{Adaptive batch size training.} Recent work on DL training algorithms have explored dynamically adapting batch sizes for better efficiency and parallelization.
AdaBatch~\cite{DBLP:journals/corr/abs-1712-02029} increases the batch size at pre-determined iterations during training, while linearly scaling the learning rate.
Smith et al.~\cite{DBLP:journals/corr/abs-1711-00489} suggest that instead of decaying the learning rate during training, the batch size should be increased instead.
CABS~\cite{DBLP:journals/corr/BallesRH16} adaptively tunes the batch size and learning rate during training using similar gradient statistics as \oursystem{}.

These works have a common assumption that extra computing resources are available to parallelize larger batch sizes whenever desired, which is rarely true inside shared-resource environments. \oursystem{} complements existing adaptive batch size strategies by adapting the batch size and learning rate in conjunction with the amount of resources currently available. Alternatively, anytime minibatch~\cite{ferdinand2018anytime} adapts the batch size to mitigate stragglers in distributed training.

KungFu~\cite{258904} supports adaptive training algorithms, including adaptive batch sizes, by allowing applications to define custom adaptation policies and enabling efficient adaptation and monitoring during training. Although KungFu is directed at single-job training and \oursystem{} at cluster scheduling, we believe KungFu offers useful tools which can be used to implement the adaptive policies used by the \oursystem{}Agent.

\noindent\textbf{Hyper-parameter tuning.} A large body of work focuses on tuning the hyper-parameters for ML and DL models~\cite{hutter2011sequential,feurer2015efficient,bergstra2011algorithms,neiswanger2019probo,kandasamy2019tuning}, which typically involves many training jobs~\cite{10.1145/3097983.3098043,katib} as discussed earlier. 
Although batch size and learning rate are within the space of hyper-parameters often optimized by these systems, \oursystem{}'s goal is fundamentally different. Whereas HPO algorithms search for the highest model quality, \oursystem{} adapts the batch size and learning rate for the most efficient execution for each job, while not degrading model quality.

\section{Conclusion}

\oursystem{} is a DL cluster scheduler that co-adaptively allocates resources, while at the same time tuning each training job to best utilize those resources. We present a formulation of goodput that combines system throughput and statistical efficiency for distributed DL training. Based on the principle of goodput maximization, \oursystem{} automatically and jointly tunes the resource allocations, batch sizes, and learning rates for DL jobs, which can be particularly difficult for users to configure manually. 
\oursystem{} outperforms and is more fair than
recent
\qirong{Is ``SOTA" too strong a claim now that new papers are out? Would it be better to (honestly) call out that we simply beat Optimus and Tiersas?} \greg{I replaced SOTA with recent here... but, do we need to do the same thing in abstract, eval and intro?  Also shortened, so take a look ;).  And, wondering if we should we drop the "fair" aspect from the conclusion.} 
DL schedulers, even if users can configure their jobs well, and provides even bigger benefits with more realistic user knowledge. 



\section{Acknowledgements}

We thank our shepherd, Michael Isard, and the anonymous OSDI reviewers for their insightful comments and suggestions that improved our work.
We also thank our colleagues from Petuum --- Omkar Pangarkar, Richard Fan, Peng Wu, Jayesh Gada, and Vishnu Vardhan --- for their invaluable contributions toward the open source implementation of \oursystem{}.

\bibliographystyle{plain}
\bibliography{references}



\end{document}